\documentclass{aa}  
\usepackage{graphicx}
\usepackage[varg]{txfonts}
\usepackage{multirow}
\usepackage{xcolor}
\usepackage{amsmath}
\newcommand{\Ha}{H$\alpha$}
\usepackage[colorlinks=true,citecolor=blue]{hyperref}

\begin{document}

   %\title{VLT/MUSE as a high-contrast imager}
   \title{Searching for proto-planets with MUSE}

   \author{C.~Xie\inst{1}
          \and 
          S.~Y.~Haffert\inst{1, 2}\thanks{NASA Hubble Fellow} 
          \and
          J.~de~Boer\inst{1}
          \and
          M.~A.~Kenworthy\inst{1}
          \and
          J.~Brinchmann\inst{1,3}
          \and
          J.~Girard\inst{4}
          \and
          I.~A.~G.~Snellen\inst{1}
          \and
          C.~U.~Keller\inst{1}
         }
   \institute{Leiden Observatory, Leiden University, PO Box 9513, 2300 RA Leiden, The Netherlands
              \email{xie@strw.leidenuniv.nl}
         \and
            Steward Observatory, 933 North Cherry Avenue, University of Arizona, Tucson, AZ 85721, USA
        \and
            Instituto de Astrof\'{i}sica e Ci\^{e}ncias do Espa\c{c}o, Universidade do Porto, CAUP, Rua das Estrelas, PT4150-762 Porto, Portugal
        \and 
            Space Telescope Science Institute, Baltimore 21218, MD, USA}

   \date{Received ---; accepted ---}

% \abstract{}{}{}{}{} 
% 5 {} token are mandatory
 
  \abstract
   {
   Protoplanetary disks contain structures such as gaps, rings, and spirals, which are thought to be produced by the interaction between the disk and embedded protoplanets. However, only a few planet candidates are found orbiting within protoplanetary disks, and most of them are being challenged as having been confused with disk features.
   }
  %However, MUSE is not designed as a high-contrast imager.
   {
   The VLT/MUSE discovery of PDS~70~c demonstrated a powerful way of searching for still-forming protoplanets by targeting accretion signatures with medium-resolution integral field spectroscopy. We aim to discover more proto-planetary candidates with MUSE, with a secondary aim of improving the high-resolution spectral differential imaging (HRSDI) technique by analyzing the instrumental residuals of MUSE.
   }
   {
   We analyzed MUSE observations of five young stars with various apparent brightnesses and spectral types. We applied the HRSDI technique to perform high-contrast imaging. The  detection limits were estimated using fake planet injections.
   }
   {
    With a 30 min integration time, MUSE can reach 5$\sigma$ detection limits in apparent \Ha~line flux down to 10$^{-14}$ and 10$^{-15}$ erg~s$^{-1}$~cm$^{-2}$ at 0.075\arcsec~and 0.25\arcsec, respectively. 
    In addition to PDS~70~b and~c, we did not detect any clear accretion signatures in PDS~70, J1850-3147, and V1094~Sco down to 0.1\arcsec.
    MUSE avoids the small sample statistics problem by measuring the noise characteristics in the spatial direction at multiple wavelengths.
    We detected two asymmetric atomic jets in HD~163296 with a very high spatial resolution (down to 8 au) and medium spectral resolution ($R \sim 2500$).
   }
  {
  The HRSDI technique when applied to MUSE data allows us to reach the photon noise limit at small separations (i.e., \textless~0.5\arcsec). With the combination of high-contrast imaging and medium spectral resolution, MUSE can achieve fainter detection limits in apparent line flux than SPHERE/ZIMPOL by a factor of $\sim$5. MUSE has some instrumental issues that limit the contrast that appear in cases with strong point sources, which can be either a spatial point source due to high Strehl observations or a spectral point source due to a high line-to-continuum ratio. We modified the HRSDI technique to better handle the instrumental artifacts and improve the detection limits. To avoid the instrumental effects altogether, we suggest faint young stars with relatively low \Ha~line-to-continuum ratio to be the most suitable targets for MUSE to search for potential protoplanets.
   }


\maketitle
%
%________________________________________________________________

\section{Introduction}

%\subsection{Imaging protoplanets and protoplanetary disks}
%\blue{\bf what makes MUSE unique?}

In the past decade many  have searched for newly forming planets (protoplanets) in protoplanetary disks, which are thought to be the birthplace of protoplanets. Recent observations at infrared (IR) and millimeter wavelengths have found protoplanetary disks containing structures such as gaps \citep{ALMAPartnership2015, Thalmann2015, Andrews2016, DSHARP_I_Andrews2018}, rings \citep{Andrews2016, DSHARP_I_Andrews2018, vanTerwisga2018, Muro-Arena2019}, and spirals \citep{Muto2012, Garufi2013, Benisty2015, Reggiani2018}. The interaction between the disk and embedded planets is one of the possible explanations for  such disk features \citep{KleyNelson2012, Pinilla2012, deJuanOvelar2013, Zhu2014, Dong2015a, Dong2015b}. Furthermore, disk morphology can be used to constrain the properties of embedded planets, such as masses and locations \citep{FungDong2015, DongFung2017}. Alternative theories were also proposed to explain the existence of disk features, such as magneto-hydrodynamic instabilities \citep{Flock2015}, and snow lines \citep{Zhang2015}. Despite all the theoretical work, the origin of the gaps and rings still remains unclear \citep{vanderMarel}. Observing protoplanets and their environments is important to understand planet-disk interaction and planet formation. 

High-contrast high spatial resolution observations have thus far found a small number of point-like sources embedded in protoplanetary disks, such as LkCa 15 b, c, and d \citep{Kraus2012, Sallum2015}; HD 100546 b and c \citep{Brittain2013, Brittain2014, Quanz2013, Quanz2015, Currie2015}; HD 169142 b \citep{Quanz2013, Biller2014}; MWC 758 \citep{Reggiani2018}; and PDS 70 b and c \citep{Keppler2018, Muller2018, Wagner2018, Haffert2019}. However, these candidate protoplanets are either being challenged or are lacking further observational confirmation \citep{Thalmann2016, Follette2017, Rameau2017, Ligi2018, Sissa2018, Huelamo2018, Currie2019, Wagner2019}, except PDS 70 b and c \citep{Keppler2018, Haffert2019, Mesa2019a}. Until the observation of  PDS 70 b and c, there had been no unambiguous evidence for a protoplanet embedded in a protoplanetary disk\footnote{The famous $\beta$ Pictoris b is orbiting in a debris disk, which is a later stage of disk evolution after the depletion of gas material in the protoplanetary disk.}.

The accretion signature used to discover PDS~70~c originates from shock-heated hydrogen gas with a temperature of $\sim$10,000 K falling in from the circumplanetary disk onto the planet \citep{Zhu2015}. Accretion shocks can excite or ionize the hydrogen atoms, which will later recombine and radiate \Ha~emission \citep{Calvet1998, Marleau2017}. Being one of the most prominent lines produced by accretion processes, the \Ha~emission line has been detected in low-mass stars \citep{Reipurth1996, Gullbring1998, Rigliaco2012}, brown dwarfs \citep{Santamaria-Miranda2018}, and the planet PDS 70 b \citep[][]{Wagner2018}. Other hydrogen accretion lines, such as H$\beta$ and Paschen $\beta$  have also been observed on substellar companions \citep{Natta2004, Rigliaco2012, Santamaria-Miranda2018}. Planetary accretion is a very energetic process where a significant amount of flux can be emitted in a very small spectral window. The accretion luminosity can even be comparable to the total internal luminosity \citep{Mordasini2017}, which enhances the contrast ratio of planet to star in \Ha~emission and makes it easier to detect. Such enhancement in contrast is one of the drivers for the search of accretion signatures. Observations of accretion signatures in \Ha~has been conducted in the past few years with MagAO \citep{Close2014, Wagner2018} and SPHERE \citep{Cugno2019, Zurlo2020}, yielding the detection of \Ha~emission from substellar companions.

Commonly used observational techniques such as angular differential imaging
\citep[ADI;][]{Marois2006_ADI} and polarimetric differential imaging \citep[PDI;][]{Kuhn2001_PDI} can generate point-like structures from asymmetrical features in the surrounding circumstellar disk \citep{Follette2017, Ligi2018}. In the worst scenario, it may lead to false identification of planets. Moreover, even if a point-like structure is not generated by aggressive post-processing algorithms, the detection of such hot spots in a disk can still be linked to local overdensities of dust. These post-processing artifacts can be sidestepped by observing at higher spectral resolution, which has an added benefit of better matching the spectral resolution of the features we try to observe. The commonly used narrowband filters of ZIMPOL and MagAO are substantially larger than the H$\alpha$ line \citep{Aoyama2019}, which leads to excess noise from the stellar continuum.

The sensitivity of a high-contrast imager can be improved when it is combined with high-dispersion spectroscopic techniques \citep{Sparks2002, Riaud2007A&A...469..355R, Snellen2015, Wang2017, Hoeijmakers2018}. The reason for the enhanced contrast is that emission from a planet has different spectral features than that of the bright host star at high spectral resolution. With the medium spectral resolution ({\sl R} $\sim$ 2500 at 656 nm), the Multi-Unit Spectroscopic Explorer (MUSE; \citealt{Bacon2010SPIE}) has the capability of distinguishing the planetary emission line from the stellar emission line \citep{Haffert2019}, avoiding the potential false positive caused by scattering light from the circumstellar disk. Therefore, the MUSE detection of the \Ha~emission line can provide   unambiguous, direct evidence of the presence of accreting protoplanets embedded in their disks.

The discovery of PDS 70 c demonstrated a powerful way of searching  still-forming protoplanets by targeting accretion signatures with medium-resolution integral field spectroscopy \citep{Haffert2019}. Previous IR observations did not find the presence of PDS 70 c. The major difficulty of identifying PDS 70 c is that its location is very close to the bright ring structure. Therefore, previous IR observations would misidentify the object as a disk structure due to its elongated appearance.

In this paper we use MUSE as a high-contrast imager with high spatial (60 -- 80 mas at 656 nm) and medium spectral resolution ({\sl R} $\sim$ 2500 at 656 nm) working at optical wavelengths to search for protoplanets. MUSE is a powerful instrument that is sensitive to faint emission lines such as those originating from protoplanets embedded in protoplanetary disks or high-velocity stellar jets. The combination of spectral resolution and high-contrast imaging can improve the contrast compared to traditional imaging with   narrow- or broadband filters (i.e., ZIMPOL), which observe at low spectral resolution. %by the factor of the square root of spectral resolution. %by the factor of the square root of spectral resolution.
The data and data reduction methods are described in Sects.~\ref{sec:obsdata} and~\ref{sec:Methods}.  In Sect.~\ref{sec:cal-issues} we discuss the instrumental issues in MUSE and provided a modified high-resolution spectral differential imaging (HRSDI) technique with a successful application on V1094~Sco. In Sect.~\ref{sec:performances} we present the performance of MUSE for imaging point sources at high contrast. The example of MUSE mapping extended stellar jets is presented in Sect.~\ref{sec:mping_jets}. We end with a discussion and conclusions in Sects.~\ref{sec:discussion} and \ref{sec:conclusions}.

\section{Data}
\label{sec:obsdata}

MUSE is medium-resolution integral field spectrograph (IFS) installed on the Very Large Telescope (VLT). MUSE is renowned for being a powerful IFS for studying galaxy formation and evolution \citep{2015A&A...575A..75B, 2016A&A...587A..98W, 2017A&A...608A...1B, 2018Natur.562..229W}, supermassive black holes \citep{2015A&A...582A..63C, 2017Natur.548..304P}, and stellar populations \citep{2016A&A...588A.148H, 2018MNRAS.473.5591K}.

In 2019, MUSE offered a new narrow-field mode (NFM) covering a field of view (FoV) of 7.5\arcsec$ \times$ 7.5\arcsec with a spatial sampling of 0.025\arcsec/pixel. Working with the GALACSI Adaptive Optics system \citep{2016SPIE.9909E..1UO, 2018SPIE10703E..02M}, the spatial resolution can reach 55 mas - 80 mas and the Strehl ratio %\footnote{Strehl ratio is the ratio of peak intensity of the observed PSF with respect to a diffraction-limited PSF without aberrations.} 
is 5\% - 20\%, depending on the actual observing condition. The spectrograph consists of 24 identical integral field units (IFU) modules, which cover a wavelength range from 480 nm to 930 nm with the spectral resolving power %\footnote{The spectral resolution R is defined as $\lambda/\Delta\lambda$, where $\Delta\lambda$ is the resolution of the spectrum at a wavelength of $\lambda$.} 
of 1740 at 480 nm and 3450 at 930 nm. Since MUSE is not designed to be a high-contrast imager, it does not have a coronagraph to suppress starlight.

To analyze the capability of MUSE on high-contrast imaging, we selected five young stars with various brightnesses and spectral types, summarized in Table~\ref{tab:targets}. For more details about the targets we refer to the following papers: \object{PDS\,70}: \cite{Mesa2019b}; \object{J1850-3147}: \cite{Biller2013}; \object{V1094\,Sco}: \cite{vanTerwisga2018}; \object{HD\,100546}: \cite{Pineda2019}; and \object{HD\,163296}: \cite{Ellerbroek2014} and \cite{Isella2019}. The details of the MUSE observations can be found in Table~\ref{tab:obs_log}. %In Figure~\ref{targets_ref_spec}, we show the spectra of our selected target stars, centered around \Ha~from 6051 to 7300 \AA. 
Due to the range of target star brightnesses, different exposure settings were chosen to optimize the observation of bright ($m_{\rm R}$ = 7 mag) and faint ($m_{\rm R}$ = 12 mag) targets.

%%%%%%%%%%%%%%%%%%%%%%%%%%%%%%%%%%%%%%%%%%%%%%%%
\begin{table*}[th!]
\caption{Targets properties}             % title of Table
\label{tab:targets}      % is used to refer this table in the text
\centering                          % used for centering table
\begin{tabular}{c c c c c c c c c c}        % centered columns (4 columns)
\hline\hline                 % inserts double horizontal lines
Name & RA & Dec & Distance & $m_{\rm R}$ & SpT &    Mass  & Age & Ref. \\    % table heading 
  & (J2000) & (J2000) & (pc) & (mag) &  & (M$_{\odot}$) & (Myr) &\\
\hline                        % inserts single horizontal line
\object{PDS\,70}   & 14 08 10.26 & -41 23 53.0 & 113.4 $\pm$ 0.5 &  11.65 &  K7IVe & 0.76 &  5.4 $\pm$ 1.0 & (1) (2) \\      % inserting body of the table
\object{J1850-3147}   & 18 50 44.52 & -31 47 50.2 & 49.5 $\pm$ 0.1 &   10.61 & K8Ve &  0.87 &  -- & (3) (4) \\
\object{V1094\,Sco}   & 16 08 36.16 & -39 23 04.2 & 153 $\pm$ 1 &   12.42  & K6 &  0.79 & 2 -- 7 & (5) (6) \\
\object{HD\,100546}   & 11 33 24.94 & -70 11 42.2 & 110.0 $\pm$ 0.6 &   8.78  & B9Vne &  2.4 & 5 -- 10 & (7) (8) \\
\object{HD\,163296}   & 17 56 21.42 & -21 57 22.5 & 101.5 $\pm$ 1.2 &   6.86  & A1Vep &  2.3 & 5 -- 7 & (9) (10) (11) \\ 
\hline%inserts single line
\end{tabular}
%\begin{tablenotes}
\tablefoot{ Distances are from \cite{GaiaCollaboration2018}.
The {\sl R}-band magnitude of HD~163296 was adopted from \cite{2008ApJ...689..513T}. Unless otherwise noted, {\sl R}-band magnitudes are taken from UCAC4 catalog \citep{Zacharias+2012}. The references for spectral type, stellar mass, and stellar age are from (1) \cite{Muller2018}; (2) \cite{PecautMamajek2016}; (3) \cite{2009AJ....137.3632L}; (4) \cite{Messina2017}; (5) \cite{2017A&A...600A..20A}; (6) \cite{vanderMarel}; (7) \cite{Guimaraes2006}; (8) \cite{vandenAncker1997}; (9) \cite{2001A&A...378..116M}; (10) \cite{Montesinos2009}; (11) \cite{Vioque2018}. The full name of J1850-3147 is \object{2MASS\,J18504448-3147472}, which is also known as \object{CD-31\,16041}}.
%\item {a.} From Planck measurements \cite{planck15}.
%\end{tablenotes}
\end{table*}  
%%%%%%%%%%%%%%%%%%%%%%%%%%%%%%%%%%%%%%%%%%%%%%%%

%%%%%%%%%%%%%%%%%%%%%%%%%%%%%%%%%%%%%%%%%%%%%%%%
\begin{table*}[th!]
\caption{Log of MUSE NFM observations}             % title of Table
\label{tab:obs_log}      % is used to refer this table in the text
\centering                          % used for centering table
\begin{tabular}{c c c c c c c c c c}        % centered columns (4 columns)
\hline\hline                 % inserts double horizontal lines
Name & Prog. ID &  Observing date & $t_{\rm DIT} \times n_{\rm DIT}$$^{a}$ & On-source time & Derotation & Frames per rot.  \\    % table heading 
  &  &  &  & (s) & (degree) & \\
\hline                        % inserts single horizontal line
PDS 70   & 60.A-9100(K) & 2018-06-20 & 300 $\times$ 6 &  1800 & 90 &  2  \\      % inserting body of the table
J1850-3147   & 60.A-9482(A) & 2018-09-06 & 30 $\times$ 6 &  180 & 90 &  3  \\    &  & 2018-09-08$^{b}$ & 30 $\times$ 6 &  180 & 90 &  3  \\      % inserting body of the table

V1094 Sco   & 0103.C-0399(A) & 2019-04-14 & 300 $\times$ 8 &  2400 & 90 &  2  \\
%   &  & 2019-06-19 & 300 $\times$ 8 &  2400 & 90 &  2 & 0.1 \\
%#HD 100546   & 0103.C-0399(A) & 2019-04-14 & 1 $\times$ 4 &  4 & 0 &  4  \\
%   &  & 2019-04-14 & 4 $\times$ 40 &  160 & 45 &  8 & 0.1 \\
HD 100546   & 0103.C-0399(A)  & 2019-04-28 & 1 $\times$ 4 &  4 & 0 &  4  \\
   &  & 2019-04-28 & 4 $\times$ 40 &  160 & 45 &  8  \\
   
HD 163296   & 0103.C-0399(A) & 2019-04-29 & 2 $\times$ 4 &  8 & 0 &  4  \\
   &  & 2019-04-29 & 8 $\times$ 40 &  320 & 45 &  8  \\
   &  & 2019-04-29 & 250 $\times$ 2 &  500 & 0 &  2  \\

\hline%inserts single line
\end{tabular}
%\begin{tablenotes}
\tablefoot{$^{(a)}$~$t_{\rm DIT}$ is exposure time per image frame and $n_{\rm DIT}$ is the number of image frames.
$^{(b)}$~We found some \Ha~residuals at 60 mas  from the star center in the data observed on Sept. 8 after we analyzed the data obtained on two nights separately. The dataset is a 3 min snapshot and has a few bad spaxels in the region of the \Ha~residuals. Therefore, we excluded the data observed on the night of Sept. 8 to minimize the influence of possible instrumental artifacts. % as we explained in Section~\ref{sec:obsdata}.
}
%\item {a.} From Planck measurements \cite{planck15}.
%\end{tablenotes}
\end{table*}  
%%%%%%%%%%%%%%%%%%%%%%%%%%%%%%%%%%%%%%%%%%%%%%%%

\section{Data reduction methods}
\label{sec:Methods}

\subsection{MUSE data reduction}
\label{sec:MUSE_data_reduction_Standard_HRSDI}

The data were calibrated and reduced using the ESO MUSE pipeline\footnote{\url{http://www.eso.org/sci/software/pipelines}}, version 2.6. The details of the  MUSE pipeline can be found in the \textsl{MUSE pipeline user manual}\footnote{\url{ftp://ftp.eso.org/pub/dfs/pipelines/instruments/muse/muse-pipeline-manual-2.6.2.pdf}} and \cite{Weilbacher2020}. The MUSE pipeline can be divided into two parts, the pre-processing calibration ({\tt muse\_scibasic}\footnote{In Sect.~\ref{sec:Methods} the text in typewriter font refers to the command line in EsoRex.}) and the post-processing calibration ({\tt muse\_scipost}). Each part is responsible for correcting different instrumental effects and performing flux calibrations. During the pre-processing calibration dark field and illumination correction are optional and can be ignored, but we found significant improvement especially for bright targets ($m_{\rm R}$ < 9 mag) after performing illumination correction. %applying the twilight correction for large scale illumination.  
Therefore, we strongly recommend using the illumination correction for high-contrast imaging applications as in this work. %\blue{Referee: provide the selection criteria for the reference star for flux calibration.}
After two stages of calibrations, all the exposures that were processed separately were combined ({\tt muse\_exp\_combine}) and ready for science use. The final data product is a 3D cube with two spatial dimensions and one spectral dimension. %Below we introduce the MUSE data reduction in more detail.

To remove the starlight halo we used the HRSDI technique, first described in \cite{Haffert2019}. %A similar technique was also used for molecule mapping $\beta$~Pictoris b \citep{Hoeijmakers2018}. %HRSDI is only sensitive to the variation in sharp spectral features, broadband differential spectral features are removed during the HRSDI process. The potential reflection of starlight from the surrounding disk and continuum emission from planets are also removed.  
Based on spectral information from every spatial pixel, HRSDI removes the stellar component in two steps: the corrections for low-order spectral effects and high-order effects. A low-order effect with a spectral resolution on the order of $R\approx50-100$ is induced by diffraction \citep{antichi2009bigre}, causing a spatial pixel (spaxel) to capture different parts of the point spread function (PSF) as a function of wavelength. Such low-order effects can be corrected by doing continuum-normalization \citep{Haffert2019}. After the correction we can create the normalized reference spectrum by averaging over all spaxels and subsequently subtracting it from all normalized spaxels in order to remove the stellar component. To further remove the residuals caused by uncalibrated instrumental effects (called high-order effects), we used principal component analysis \citep[PCA;][]{Soummer2012, Amara2012}. Each PCA component corrects for different instrumental artifacts. For example, a small offset in the wavelength solution will lead to residuals with the shape of P Cygni profiles after the subtraction of the reference spectrum, which could be corrected by the first few PCA components. The residual effects of the telluric lines could also be corrected by the first few PCA components. The number of components to subtract is determined by maximizing the signal-to-noise ratio (S/N) of injected fake planets. %Recently, \cite{Hashimoto2020} introduced a slightly different approach based on data of PDS70. They combined the two steps of correction and used PCA to create a normalized reference spectrum to remove the stellar component. While we only used PCA to remove the high-order residuals.
In Sect.~\ref{sec:cal-issues}, we provide a modified HRSDI technique to minimize instrumental issues in MUSE by adjusting  how we build the reference spectra.

\subsection{Fake planet injection}
\label{sec:planet_inj}

\begin{figure}
\centering
\includegraphics[width=0.48\textwidth]{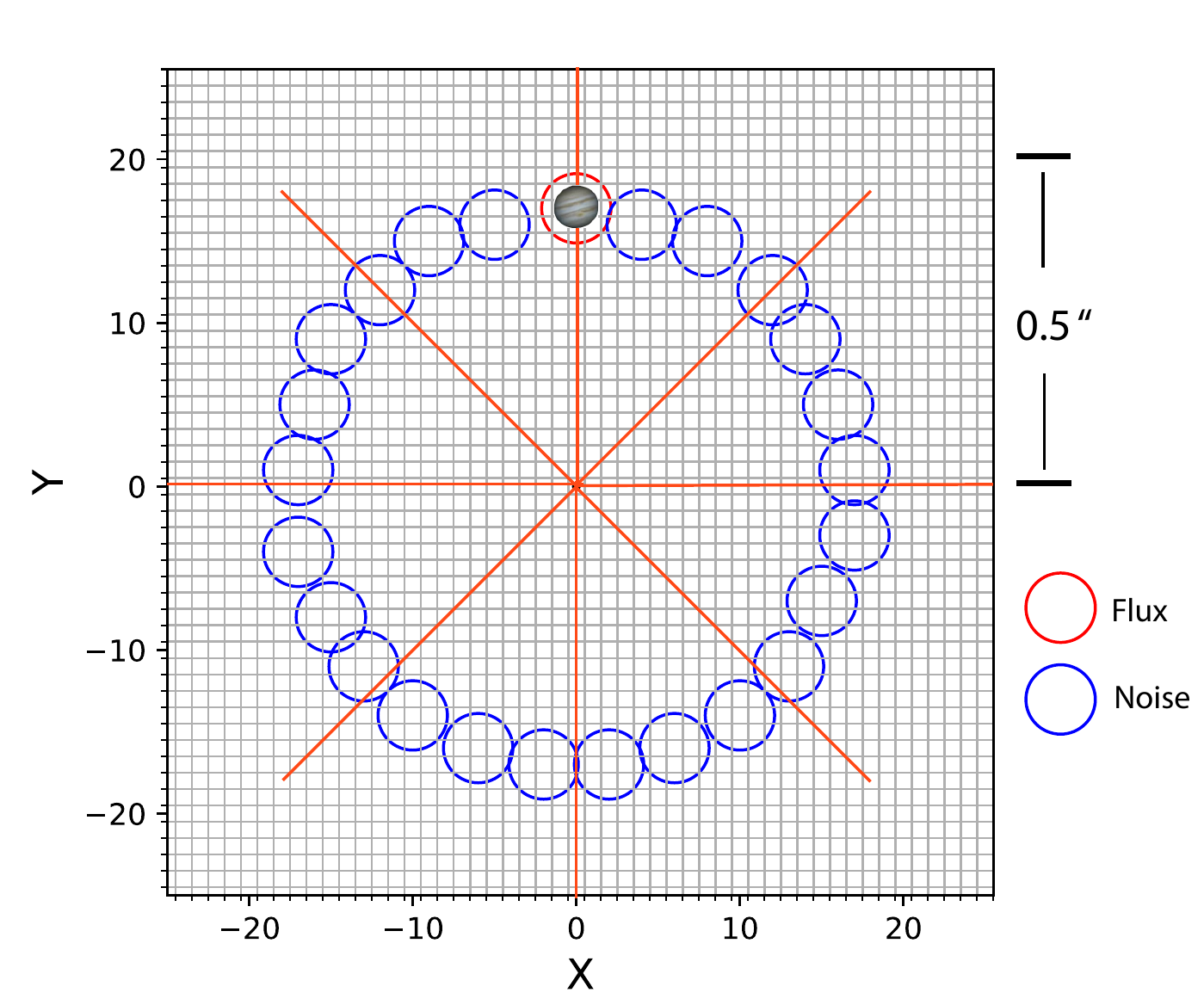}
%\vspace{-8mm}
\caption{Illustration of our fake planet injection. A fake planet (red circle) is injected in one of eight position angles (red lines) in a spacing of 1 spaxel. The blue circles represent the regions where the corresponding noise is measured. }
\label{inj_rms}
\end{figure}

To quantify the performance of any high-contrast imager, a standard approach is to inject fake planet signals and then recover them to estimate the sensitivity. As illustrated in Fig.~\ref{inj_rms}, we used eight different position angles (PAs; separated by 45$^{\circ}$) where the artificial planet was injected with varying contrast levels. The radial dependence was estimated by injected planets by increasing the radial separation in steps of 1 spaxel. The artificial planet was created based on a planet spectrum and the stellar PSF. The planet spectrum was taken as a single Gaussian line at a specific wavelength of our interest, with a full width at half maximum (FWHM) of $\sim$135 km/s. Such a line profile is covered by three spectral channels. %The line width is similar to the actual accreting planet. 
For each spectral channel we estimated the planet PSF by taking the corresponding observed stellar PSF and normalized it by the total flux in the FoV. After shifting the PSF and scaling the contrast, only one planet was injected for each of the fake planet injection tests. It is possible to inject multiple planets at the same time to speed up the calculations, but we opted to inject a single planet at a time for simplicity.

The noise was estimated by performing aperture photometry in the spectral channels of our interest after processing the data with HRSDI. Figure~\ref{inj_rms} shows the regions where we injected a fake planet (red circle) and the regions we used to estimate the noise (blue circles). Inside each circle a square aperture of 3 by 3 spaxels (75 by 75 mas) was adopted to match the resolution of the data, which is $\sim$80 mas at \Ha. The adopted size of apertures depends on the actual FWHM of the dataset. Due to the low Strehl ratio (\textless~20\%), the aperture with a size of the FWHM has significant flux loss. This flux loss was corrected by dividing by the encircled energy in the aperture, which was measured on the star by taking the ratio of flux within the aperture and within 1\arcsec~radius. The flux correction due to the PSF subtraction was automatically done for fake planets because the injected flux was known when the recovered flux reached given S/N.

Different contrast values were chosen to derive the relation between the contrast and the S/N. The contrast ratio is defined as the ratio between the injected line flux of the planet compared to the stellar flux (continuum + line) at the same wavelength. After that, we derived a contrast curve at given S/N (5$\sigma$). The contrast value was given in the unit of magnitudes.
%, which is calculated by
%\begin{equation}
%\Delta Mag = -2.5log_\textrm{10}(Flux_\textrm{planet}/Flux_\textrm{star}).
%\end{equation}
The uncertainty of the contrast curve was estimated by calculating the standard deviation of 5$\sigma$ contrasts obtained at four or eight PAs, depending on the separation bin. 

%%%%%%%%%%%%%%%%%%%%%%%%%%%%%%%%%%%%%%%%%%%%%%%%%%%%%%%%%%%%%%%%%%%%%%%%%%%%%%%%%%%%%%%%%%%%%%%%%%%%%%

\section{Instrumental issues for strong \texorpdfstring{\Ha}~emitter}
\label{sec:cal-issues}
\subsection{Variation in \texorpdfstring{\Ha}~line-to-continuum ratio.}

\begin{figure}
   \centering
    \includegraphics[width=0.49\textwidth]{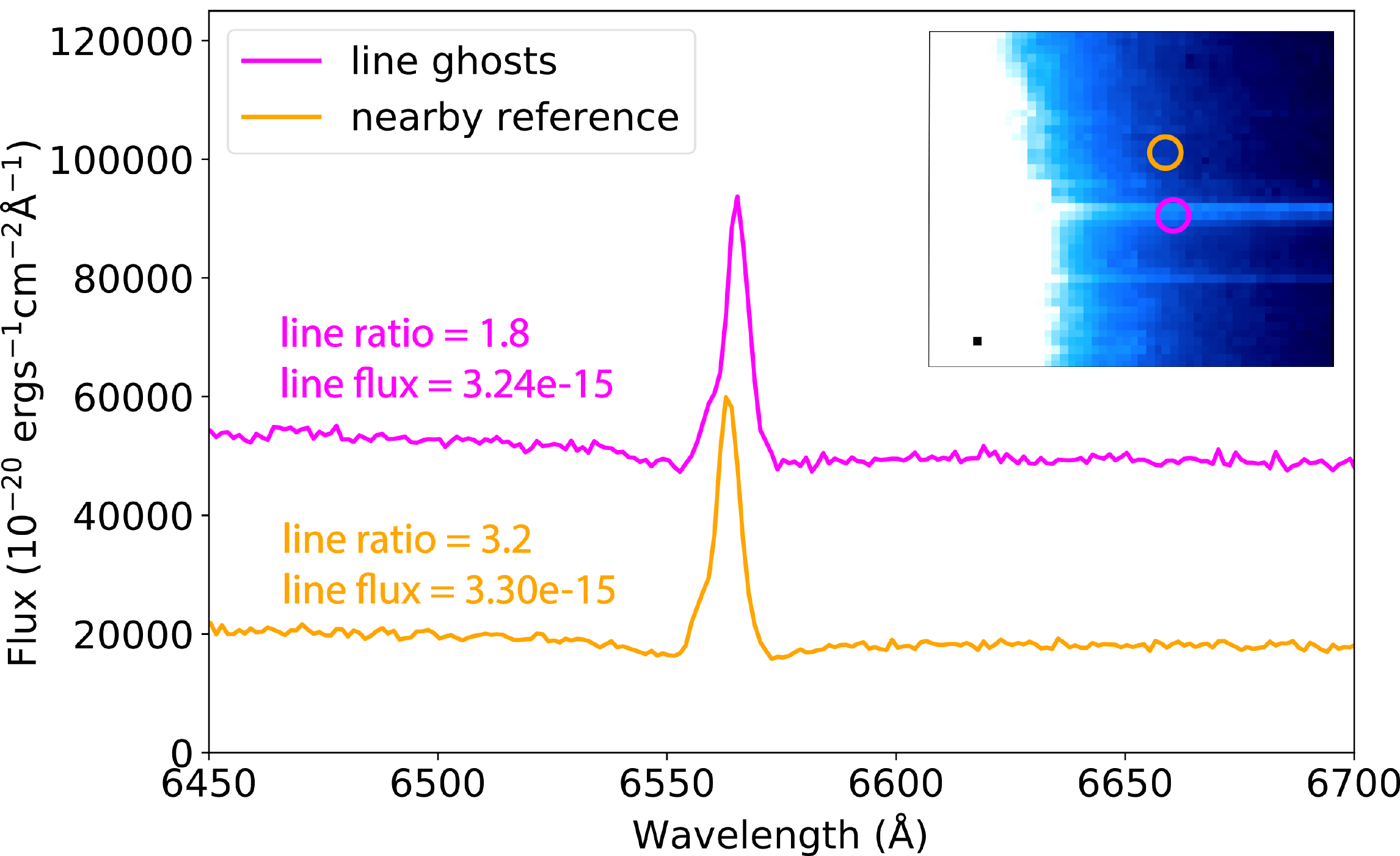}
      \caption{Illustration of the effect caused by line ghosts. The spectrum of the lines ghost is similar to the original spectrum with a lift of additional flux. The inset   shows line ghosts in the zoomed-in image of HD~163296 at 6564~\AA~and the regions (orange and magenta circles) where  the corresponding spectra were extracted. The line ratio is defined as peak line flux over the local continuum. The line flux was measured by fitting simple Gaussian profile with continuum removed. The line flux is in units of erg~s$^{-1}$~cm$^{-2}$. % The star is located at the left side of the image.
      }
         \label{line_ghosts}
\end{figure}

MUSE was not designed for high-contrast imaging. During our analysis, we found that MUSE has an instrumental artifact that causes the stellar \Ha~line-to-continuum ratio to vary across the field. The residuals due to the instrumental effects can be stronger than those from photon noise, prohibiting us from reaching the photon noise detection limits at \Ha~(see Table~\ref{tab:noise_fit_results} and Sect.~\ref{sec:noise_component} for the details about noise decomposition). Moreover, the objects with either high Strehl or with high \Ha~line-to-continuum ratios (see Table~\ref{tab:target_flux} for V1094~Sco, HD~163296, and HD~100546) have significant variations at \Ha~that exceed the noise at nearby wavelengths. 

The variations show imprints of the IFU configuration, indicating an instrumental origin. For sources with strong stellar \Ha\ emission, this variation can be larger than the \Ha\ emission from a putative planet. As a result, the MUSE performance as a high-contrast imager is limited around such strong \Ha~emission line. Nevertheless, for other wavelengths without strong line emission of the star, MUSE can reach the photon noise detection limit (e.g., detecting stellar jets shown in Sect.~\ref{sec:mping_jets}).

The reasons for the non-uniform line-to-continuum ratios across the field are still unknown. One possible explanation is ghosts, identified as an instrumental artifact in \cite{Weilbacher2015}. Hereafter we refer to the ghosts identified in \cite{Weilbacher2015} as line ghosts in order to distinguish them from the blob ghost we found in HD~163296 (see also Fig.~\ref{fig:blob_ghost} and Appendix \ref{sec:blob_ghost} for  details).  

In our data line-to-continuum ratios are much lower at the location of line ghosts compared to nearby spaxels where line ghosts are less prominent (see Fig.~\ref{line_ghosts}). The change in line-to-continuum ratio is similar to extra background light adding on to the original spectrum. Line ghosts are present in all of our targets, visible in the stellar images in the form of overbright strips at all wavelengths. From the perspective of high-contrast imaging, although line ghosts are very prominent at the IFU in which the PSF core is located, it should be noted that line ghosts exist across the entire field at different levels. 

Another explanation is the presence of spatio-spectral cross-talk in the spectral extraction step. We see that the effect is stronger when there is a strong point source, either spatially because of high Strehl or spectrally in the case of the high \Ha-to-continuum ratio, in the data. When a strong point source is present the wings of the line spread function (LSF) can overflow into adjacent spectral and spatial bins. This can be calibrated if the LSFs during the observations are known. The MUSE data processing pipeline takes the LSF into account during the spectral extraction by using LSF profiles measured during the day, but due to non-common path errors the LSF can actually change during the observations, which may lead to small errors in the extracted spectra. 

\begin{figure*}
   \includegraphics[width=\textwidth]{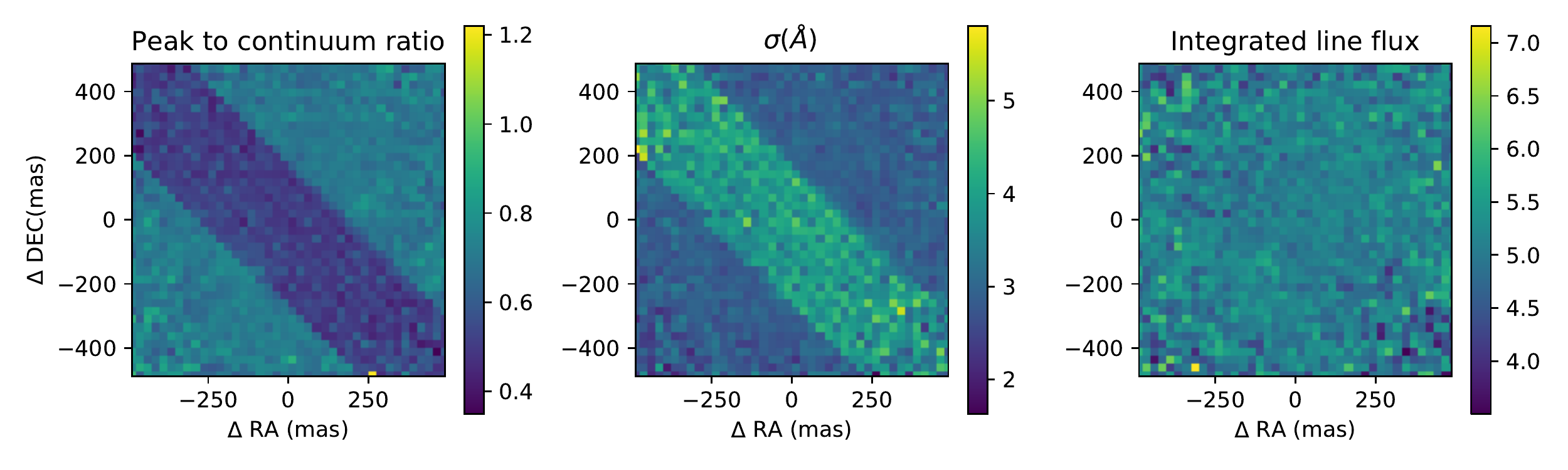}
  \caption{One of most affected datacubes from the observations of PDS70. The broad band is aligned with the slicer orientation if transformed to x-y instrument coordinates. The left figure shows the peak-to-continuum ratio. The middle figure shows the measured line width. There is a strong negative correlation between the line width and the peak to continuum ratio. The figure on the right shows the total line integrated flux, which is constant over the field. This indicates that the LSF is varying over the field.}
   \label{fig:crosstalk}
\end{figure*}

By analyzing the power in the H$\alpha$ we  found evidence in favor of spatio-spectral cross-talk as opposed to the line ghost explanation. For each spaxel we fitted the following profile,
\begin{equation}
    \phi = b \left(1 + a \exp{\left[\left(\lambda-\mu\right)^2/\sigma^2\right]}\right).
\end{equation}
Here $b$ is the local continuum, and $a$ is the line peak to continuum ratio. The line position, $\mu$, and width, $\sigma$, were fitted at the same time. The results of the fit for each spaxel can be seen in Fig.~\ref{fig:crosstalk}. There seems to be a strong correlation between the variations in the peak-to-continuum ratio and the line width. The correlation is expected if the LSF is broadening across the field. The total integrated power of the fitted Gaussian line is $P=a\sqrt{\pi \sigma^2}$. The total integrated power is shown in the third panel of Fig.~\ref{fig:crosstalk}. From this we can see that the integrated power is constant over the field, indicating that the LSF is varying over the field instead of a ghost. The variation also seems to be aligned with the slicer orientation, which indicates to instrumental effects. If this is the origin of the instrumental cross-talk it would be very difficult to correct as a full diffractive model of the whole instrument would need to be modeled for every exposure. Future European Extremely Large Telescope instruments that also include image slicers, such as HARMONI and METIS, should take these effects into account for their sensitivity estimates of their high-contrast imaging modes.

%%%%%%%%%%%%%%%%%%%%%%%%%%%%%%%%%%%%%%%%%%%%%%%%%%%%%%%%%%%%%%%%%%%%%%%%%%%%%%%%%%%%%%%%%%
%%%%%%%%%%%%%%%%%%%%%%%%%%%%%%%%%%%%%%%%%%%%%%%%%%%%%%%%%%%%%%%%%%%%%%%%%%%%%%%%%%%%%%%%%%

\subsection{Modified HRSDI}
\label{sec:modified_HRSDI}

The instrumental residuals of MUSE have variations on spatial scales comparable to the FWHM across the FoV, thus solving this issue is far beyond the scope of this paper. However, we attempt to minimize the effect during the removal of stellar emission. Below we propose modified HRSDI aiming to build better reference spectra to reduce the impact of the instrumental effects. In addition, we also recommend an extra wavelength calibration to further correct the wavelength offset without involving PCA.

\subsubsection{Building reference spectra annularly}
\label{sec:annularly}

Modified HRSDI only adjusts the way of building the reference spectrum, compared with standard HRSDI. Inaccurate reference stellar spectra would lead to strong residuals after subtracting the reference spectrum from observed data at every spaxel. Although such residuals can be removed through the latter step with PCA, this might also reduce emission from potential planets at the same wavelength. In the worst case scenario the residuals have similar spectral features to planets, and aggressive PCA could remove the planet emission. Nevertheless, the aggressive PCA can be avoided if the reference spectrum matches the observed stellar spectrum for every spatial pixel as accurately as possible. 

To build reference spectra that better match observed data, we propose to build the reference spectrum annularly, instead of only building one reference spectrum as in standard HRSDI. The assumption is that the line-to-continuum ratios are relatively uniform at the same separations. %\LEt{ at some separations? } %In such a case, different exposures with multiple position angles were first combined and then processed with modified HRSDI.
The size of the annuli was determined by minimizing the residuals after the subtraction, but are still large enough to reduce the influence of potential planet spectra on reference spectra.

\subsubsection{Extra wavelength calibration}
\label{sec:extra_wavelength_cal}

The temperature difference between nighttime science observations and daily calibrations will cause offsets in the wavelength solution. Without considering the temperature difference (i.e., exclusively used the arc frames from daily calibrations), the accuracy of the wavelength calibration is not greater than 0.4~\AA. Two sky lines (5577.339 \AA~and 6300.304 \AA)  obtained with nighttime science exposures are used to correct for such offsets caused by the temperature difference in the instrument. However, the accurate measurement of sky lines requires a sufficient integration time. For integration times longer than 100~s, the accuracy of the wavelength calibration can be improved to about $\sim$0.1~\AA\footnote{\url{https://www.eso.org/sci/facilities/paranal/instruments/muse/doc.html}}.

\begin{figure*}
   \centering
    \includegraphics[width=0.95\textwidth]{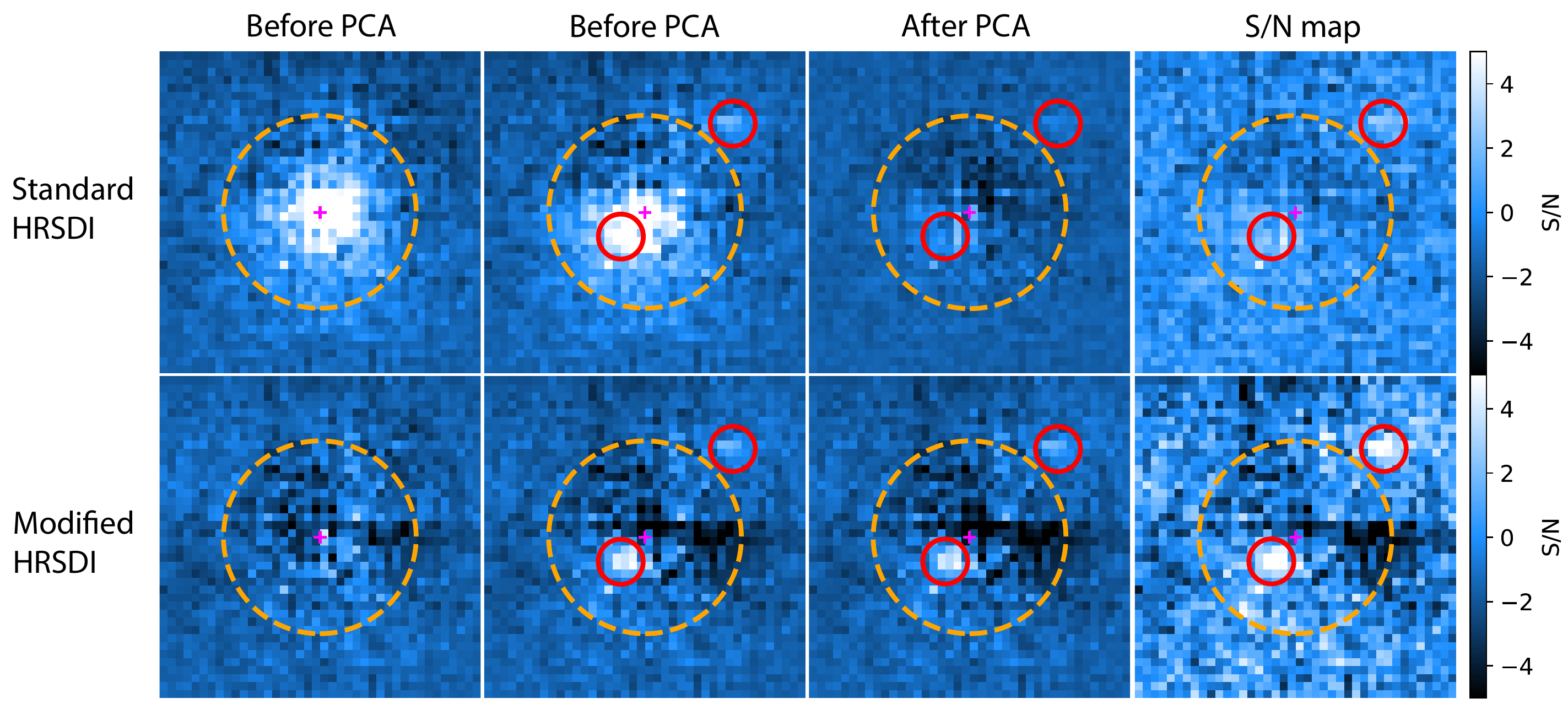}
      \caption{Comparison \Ha~images of V1094 Sco  processed in standard HRSDI (top panel) and modified HRSDI (bottom panel). The first three columns have the same color scales. Column 1: Residual images after the low-order correction in HRSDI. Column 2:  Same image as  Col. 1, but with two injected fake planets at the radius of 0.1\arcsec and 0.35\arcsec (red circles). Column 3: Residual images after the high-order correction (PCA). Column 4: Corresponding S/N maps. The orange dashed circle is of radius r = 0.3\arcsec, and the magenta cross indicates the star center. The FWHM of the V1094 Sco dataset is $\sim$150 mas (6 spaxels across). %\red{\textbf{For displaying purpose, all eight images were smoothed with a Gaussian filter of $\sigma$ = 1.}} 
      Compared with standard HRSDI, modified HRSDI can effectively remove stellar emission after low-order correction and recover most of planet emission down to 0.1\arcsec.}
         \label{mod_HRSDI_vs_stand_HRSDI}
\end{figure*}

The offset of the wavelength calibration can be corrected by the first few PCA components. However, as shown in Fig.~\ref{mod_HRSDI_vs_stand_HRSDI}, PCA will also reduce planet emission in standard HRSDI if the data has instrumental issues. Performing an additional wavelength calibration is  useful if we aim to reduce the known instrument effect without performing PCA.%\LEt{ =1st conditional, likely (... could be useful if our aim were to reduce=2nd conditional, hypothetical) } 
By shifting the observed spectra with respected to the reference spectrum, an offset map of wavelength calibration can be calculated by performing the chi-squared test at every spaxel. The stellar \Ha~emission line was used to find the remaining offsets in the wavelength solution. In the case of V1094~Sco, which  has long  integration time (300~s), the average offset is around 0.1~\AA~or less at the center high S/N region. Instead,  HD~163296 was observed with short integration time (\textless 10~s), and the offsets can be larger than 0.3~\AA. 

\begin{figure}
\centering
\includegraphics[width=0.49\textwidth]{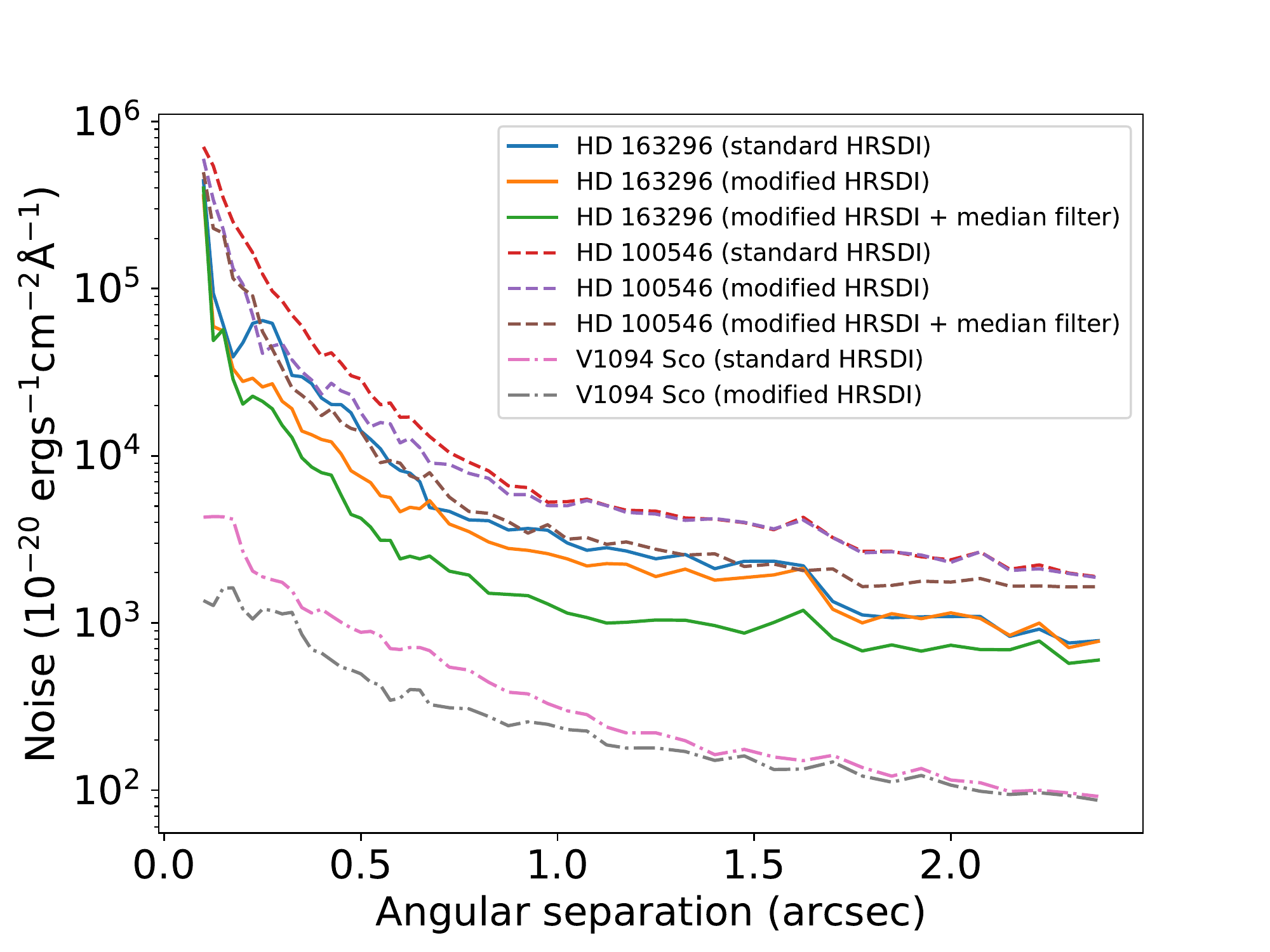}
%\vspace{-8mm}
\caption{Noise in residual images as a function of angular separations. Compared to standard HRSDI, modified HRSDI built multiple reference spectra annularly (see Sect.~\ref{sec:annularly} for details) and applied extra wavelength calibration (see Sect.~\ref{sec:extra_wavelength_cal} for details). Additional median filters were applied on the residual images processed by modified HRSDI. No PCA was used to avoid reducing planet emission caused by aggressive PCA.}
\label{improvement_in_noise}
\end{figure}

\subsection{Application: V1094 Sco}

The MUSE data of V1094~Sco has non-uniform line-to-continuum ratios across the field. %V1094 Sco has a medium line-to-continuum ratio of 2.47, suggesting the potential instrument issues. 
Standard HRSDI leads to strong residuals after low-order correction and aggressive PCA that reduces $\sim$70\% to 100\% emission from the injected planet (see top panel in Fig.~\ref{mod_HRSDI_vs_stand_HRSDI}), depending on the injected wavelengths and locations. These residuals are not due to the offset of wavelength calibration, but are caused by the variation in the  line-to-continuum ratio. Therefore, we adopted the modified HRSDI, which builds the normalized reference spectra annularly to minimize the instrumental effects, and applied the extra wavelength calibration described in Sect.~\ref{sec:extra_wavelength_cal}.

In the application of V1094 Sco, modified HRSDI can minimize instrumental issues and recover injected planet emission down to 0.1\arcsec. %The detection limit of apparent \Ha~line flux for V1094 Sco is shown in Figure~\ref{line_flux_Macc}. \red{CX: need to revise}
Figure~\ref{mod_HRSDI_vs_stand_HRSDI} shows comparison images of V1094 Sco being processed in standard HRSDI (top panel) and modified HRSDI (bottom panel). After the low-order correction that removes broadband continuum emission, standard HRSDI failed to remove part of stellar emission within a radius of 0.3\arcsec~due to the variation in the line-to-continuum ratio. On the contrary, modified HRSDI can effectively remove stellar emission. To better characterize the performance of two methods, two identical fake planets were injected at each location, indicated by the red circles in Fig.~\ref{mod_HRSDI_vs_stand_HRSDI}. Both methods can recover the injected planet at a radius of 0.35\arcsec~if no PCA is used. The strong residual left by standard HRSDI results in aggressive PCA, which will significantly reduce planet emission. As for modified HRSDI, PCA did not remove significant contributions since the reference spectra and observed spectra better matched. Figure~\ref{improvement_in_noise} shows the improvement in modified HRSDI compared to the standard HRSDI in terms of better removing the noise by a factor of 2--3 at separations less than 0.5\arcsec.

\subsection{The case for high line ratios: HD~100546 and HD~163296}

\begin{figure*}
\centering
\includegraphics[width=0.98\textwidth]{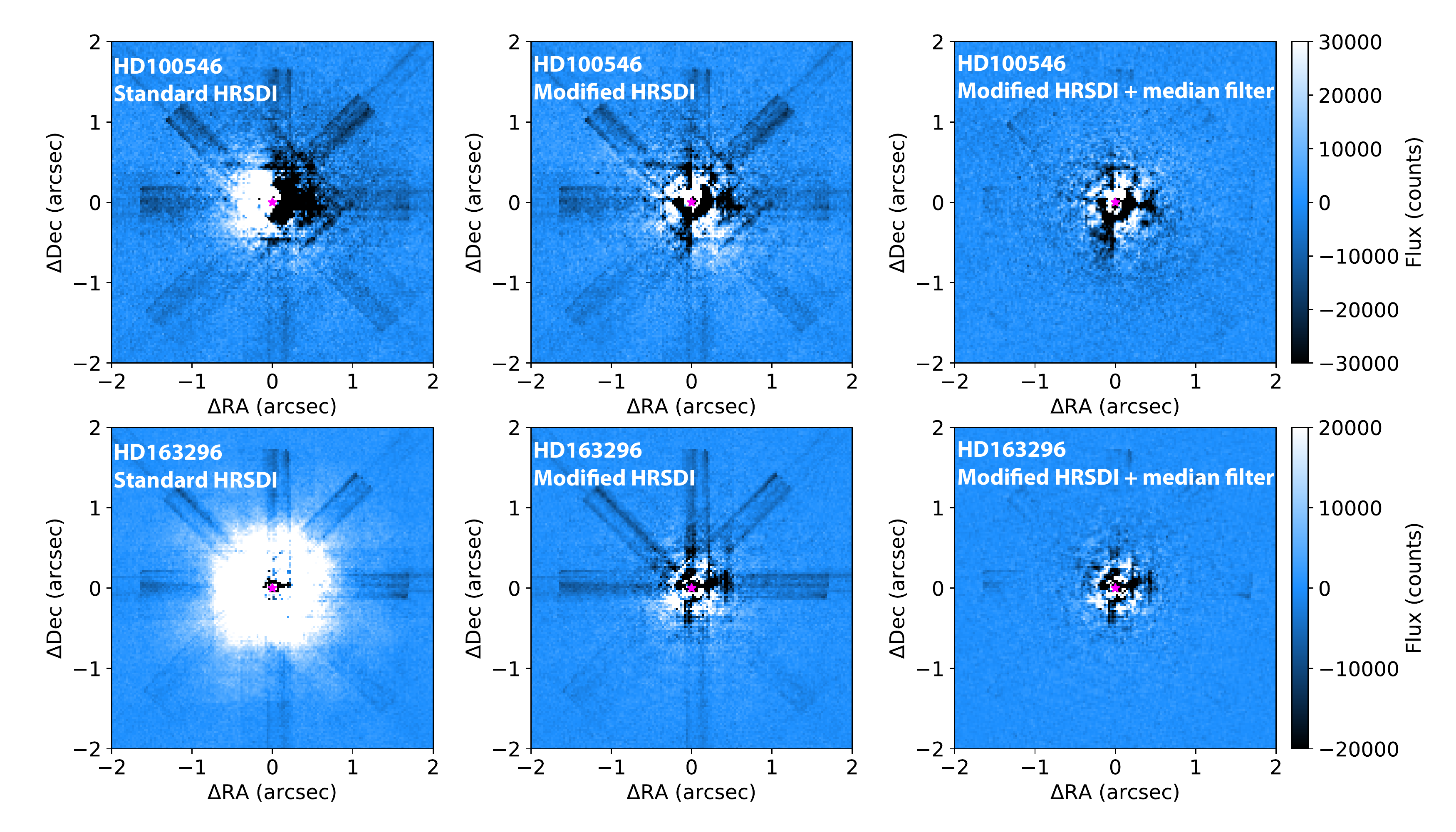}
%\vspace{-8mm}
\caption{Residual images of HD~100546 and HD~163296 at \Ha~processed by standard and modified HRSDI. No PCA was used during the post-processing. Additional median filters were applied on the residual images after the process of modified HRSDI to remove the strips due to the instrumental issues, which are shown in the third column. %Only the low-order effects and wavelength offset were corrected \red{\textbf{during the process of HRSDI}}.
}
\label{HD100546_HD163296}
\end{figure*}

Figure~\ref{HD100546_HD163296} shows the residual images of HD~100546 and HD~163296 at \Ha~processed by standard and modified HRSDI. No PCA was used during the post-processing. HD~100546 and HD~163296 have very high \Ha~line-to-continuum ratios and strong continuum emission  (see Table~\ref{tab:target_flux}). Both targets show strong variations of line-to-continuum ratio across the field, resulting in the strong stellar residuals and strips with 45\textdegree~angular separations associated with 45\textdegree~derotation. To remove the strip we applied the median filter with the width of 1 pixel and length of 50 pixels along the strip direction on the residual image after the process of modified HRSDI (see Col. 3 in Fig.~\ref{HD100546_HD163296}). The quantitative improvement of each process can be found in Fig.~\ref{improvement_in_noise}. Applying the median filter can effectively remove the strips and reduce the noise at separations larger than 0.5\arcsec, while less effectively at the inner region (\textless 0.3\arcsec) where the variations become much larger.

Even with modified HRSDI there are still strong residuals due to the instrumental issues at the inner region (\textless 0.5\arcsec). Nevertheless, modified HRSDI already removed more stellar emission than standard HRSDI, as in the case of V1094~Sco in Fig.~\ref{mod_HRSDI_vs_stand_HRSDI}. Further investigation is needed to reduce the instrumental issues either via data reduction or through instrumental perspective.

%%%%%%%%%%%%%%%%%%%%%%%%%%%%%%%%%%%%%%%%%%%%%%%%%%%

\section{Point source sensitivity of MUSE}
\label{sec:performances}

\subsection{Noise components in MUSE}
\label{sec:noise_component}

In general, the noise in the high-contrast image can be written as the combination of several noise components:
\begin{equation}
\label{equ_noise_component}
\sigma = \sqrt{\sigma_{\textrm{photon}}^{2} + \sigma_{\textrm{bg}}^{2} + ...}~.
\end{equation}
Here the photon noise $\sigma_{\textrm{photon}}$ = (c$_\lambda \cdot F^{\alpha}_\lambda$) with coefficient c$_\lambda$, exponent $\alpha$, and stellar flux $F$. The background noise $\sigma_{\textrm{bg}}$ consists of the noise components that remains constant per spectral channel in the entire FoV. All three parameters, c, $\alpha$, and $\sigma_{\textrm{bg}}$, are wavelength dependent.

%%%%%%%%%%%%%%%%%%%%
\begin{figure}
\centering
\includegraphics[width=0.48\textwidth]{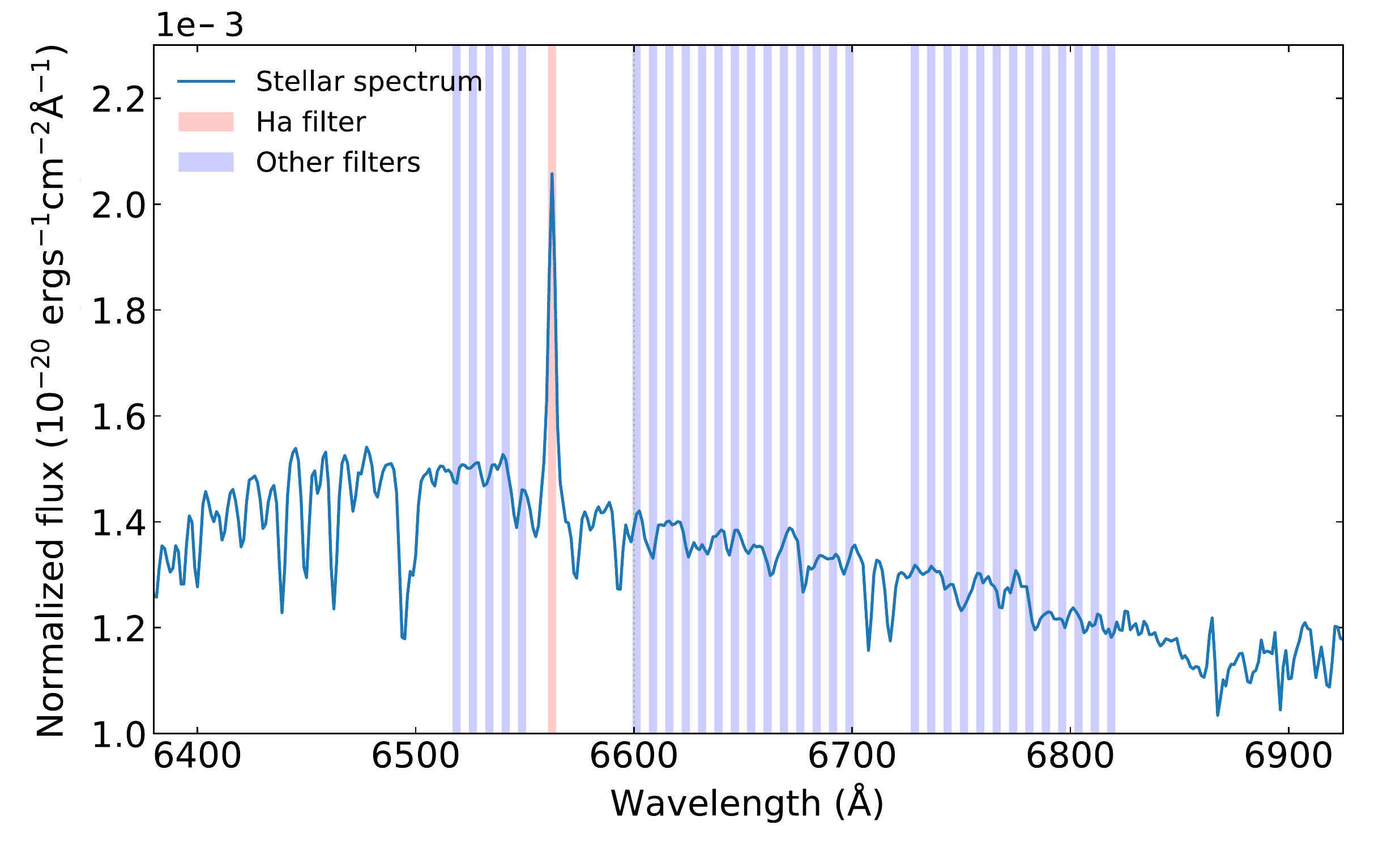}
%\vspace{-8mm}
\caption{Stellar spectrum of J1850-3147 centered around \Ha. Colored stripes indicate the filters adopted to measure the noise and stellar flux at corresponding wavelengths. Each filter contains three spectral channels of 3.75~\AA. }
\label{filters}
\end{figure}
%%%%%%%%%%

To measure the noise and stellar flux at multiple wavelengths, we first create filters around the \Ha, from 6510 \AA~to 6825 \AA~(see Fig.~\ref{filters}). 
The bandwidth of each filter is 3~$\times$~1.25 \AA, which corresponds to the FWHM of the \Ha~line. 
In each filter, we measure the noise and stellar flux by performing the aperture photometry described in Sect.~\ref{sec:planet_inj} without injecting any planets. The filters were placed with the interval of 3.75 \AA. We exclude the filters very close to \Ha~and absorption features manually by visual inspection to only measure continuum emission of the host star. %The bandwidth of each filter is 3.75 \AA (3 spectrum channels), which is around FWHM of possible \Ha line emission of planet. 

\begin{figure}
\centering
\includegraphics[width=0.48\textwidth]{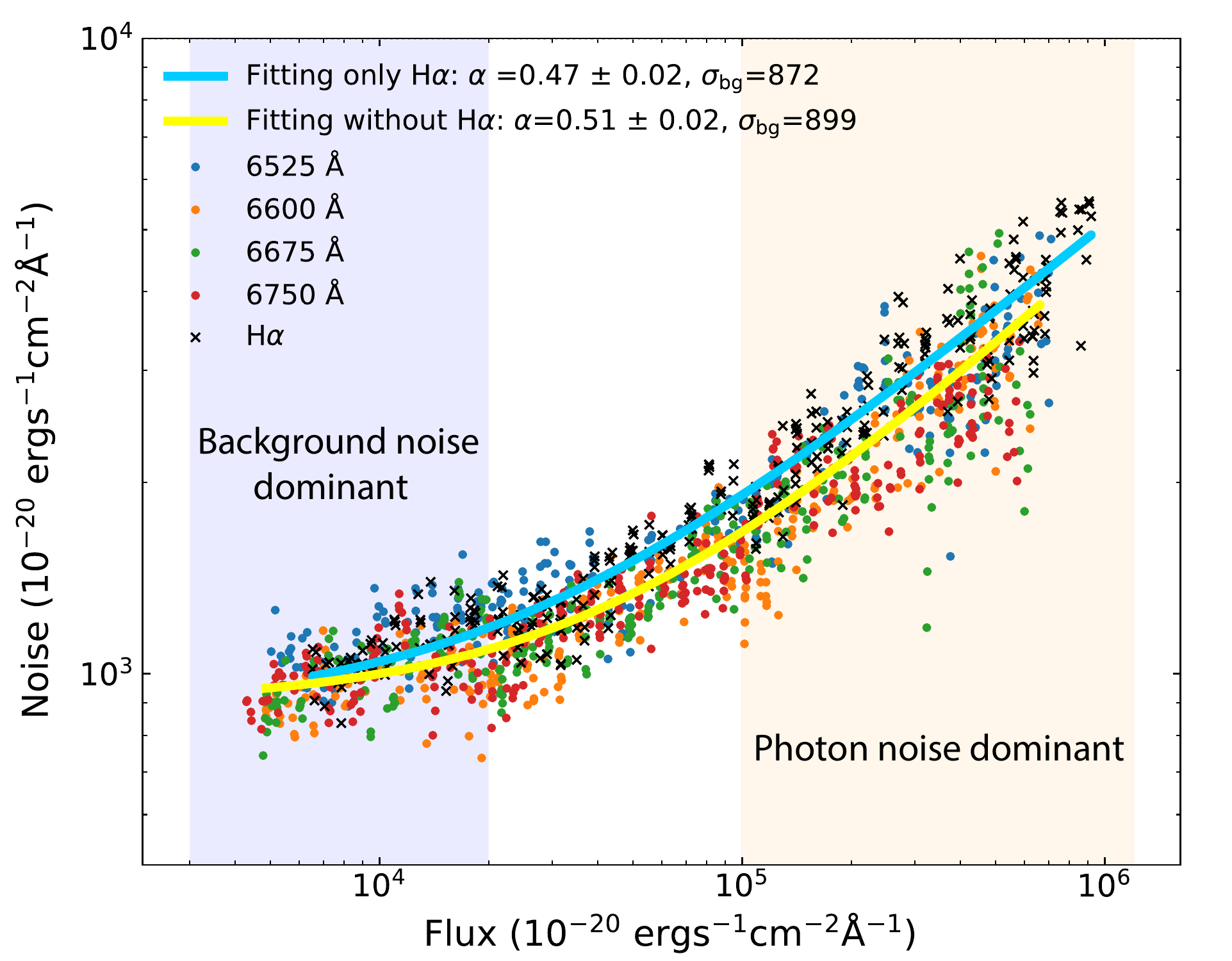}
%\vspace{-8mm}
\caption{Noise curves of J1850-3147 measured from the residual image at multiple wavelengths as a function of stellar flux obtained before performing HRSDI. Fittings of two noise components  are shown as  cyan (fitting only with \Ha~data) and yellow lines (fitting without \Ha~data). %The noise curve of the residual image at \Ha~is consistent with that of other wavelengths, which 
The noise curves are close to the theoretical limit of photon noise at small separations (\textless~0.6\arcsec; yellow shaded region). 
The fitting results are presented in Table~\ref{tab:noise_fit_results}.}
\label{counts_vs_rms}
\end{figure}

The relations between stellar flux and image noise (after HRSDI) at multiple wavelengths are presented in Fig.~\ref{counts_vs_rms}. To decompose the noise, we perform the fitting with only the photon noise and background noise, with free parameters of c$_\lambda$, $\alpha$, and $\sigma_{\textrm{bg}}$, as in Equation~\eqref{equ_noise_component}.

In the case of J1850-3147 we found the exponent $\alpha$ to be 0.47 $\pm$ 0.02 and 0.51 $\pm$ 0.02 for \Ha~only and for other wavelengths except \Ha, respectively. Both are very close to the ideal photon noise. The uncertainty of the fitting result is estimated from 1000 times bootstrap resampling. Since the stellar flux follows the PSF profile as a function of radial separations, the photon noise also follows the PSF profile. The result of fitting two noise components   suggests that MUSE is photon noise ($\sigma_{\textrm{photon}}$) dominated at small separations (\textless~0.6\arcsec) and background noise dominated at large separations (\textgreater~1.2\arcsec). The size of the region where dominated by photon noise is determined by the target brightness and background noise. In our target sample, MUSE can be considered  photon noise dominant up to the separation that ranges from 0.5\arcsec\ to 1.5\arcsec.

We present the results of two-component fitting for all targets in Table~\ref{tab:noise_fit_results}. The noise in the residual images of J1850-3147 and PDS 70 is very close to the theoretical limit of ideal photon noise after the process of standard HRSDI described in Sect.~\ref{sec:MUSE_data_reduction_Standard_HRSDI}. However, for the rest of the targets the photon noise limit can only be achieved outside the spectral window of stellar \Ha~emission (see Table~\ref{tab:noise_fit_results}). It is caused by instrumental issues in MUSE, which are described in Sect.~\ref{sec:cal-issues}.

%%%%%%%%%%%%%%%%%%%%%%%%%%%%%%%%%%%%%%%%%%%%%%%%
\begin{table*}[th!]
\caption{Noise fitting results of the target sample}             % title of Table
\label{tab:noise_fit_results}      % is used to refer this table in the text
\centering                          % used for centering table
\begin{tabular}{c c c c c c c}        % centered columns (4 columns)
\hline\hline                 % inserts double horizontal lines
Name & $c_{\textrm H\alpha}$ & $\alpha_{\textrm H\alpha}$ & $\sigma_{\textrm{bg, H}\alpha}$ & $c_{\lambda}$ & $\alpha_{\lambda}$ & $\sigma_{\textrm{bg}, \lambda}$  \\    % table heading 
\hline                        % inserts single horizontal line
PDS70   & 1.90 $\pm$ 0.56 & 0.50 $\pm$ 0.02 & 116 $\pm$ 12 &  2.37 $\pm$ 0.55 & 0.48 $\pm$ 0.02 & 128 $\pm$ 7 \\
J1850-3147   & 7.43 $\pm$ 2.05 & 0.47 $\pm$ 0.02 & 872 $\pm$ 49 &  4.10 $\pm$ 1.03 & 0.51 $\pm$ 0.02 &  899 $\pm$ 21 \\      % inserting body of the table
V1094 Sco   & 0.08 $\pm$ 0.03 & 0.70 $\pm$ 0.03 & 61 $\pm$ 2 &  1.28 $\pm$ 0.22 & 0.48 $\pm$ 0.02 & 61 $\pm$ 2 \\
%V1094 Sco$^{a}$   & 0.74 $\pm$ 0.28 & 0.54 $\pm$ 0.3 & 52 $\pm$ 9 &  1.11 $\pm$ 0.21 & 0.48 $\pm$ 0.02 & 56 $\pm$ 2 \\
V1094 Sco$^{a}$   & 0.56 $\pm$ 0.17 & 0.60 $\pm$ 0.3 & 96 $\pm$ 5 &  1.26 $\pm$ 0.21 & 0.48 $\pm$ 0.02 & 61 $\pm$ 2 \\
HD 100546   & 0.07 $\pm$ 0.06 & 0.70 $\pm$ 0.05 & 2362 $\pm$  1747 &  7.06 $\pm$ 1.70 & 0.50 $\pm$ 0.02 & 4651 $\pm$ 116 \\
HD 163296   & 0.14 $\pm$ 0.06 & 0.63 $\pm$ 0.03 & 462 $\pm$ 558 &  1.96 $\pm$ 0.49 & 0.54 $\pm$ 0.02 & 1856 $\pm$ 43 \\ 
\hline%inserts single line
\end{tabular}
%\begin{tablenotes}

\tablefoot{We adopt the same notations as in Equations~\eqref{equ_noise_component} and \eqref{equ_noise_Ha}, where photon noise $\sigma_{\textrm{photon}}$ is written as (c$_\lambda \cdot F^{\alpha}_\lambda$) and background noise is $\sigma_{\textrm{bg}}$. Unless otherwise noted, the data was processed with standard HRSDI described in Sect.~\ref{sec:MUSE_data_reduction_Standard_HRSDI}. $^{(a)}$ The data was processed in modified HRSDI (see Sect.~\ref{sec:cal-issues} for details).}% $^{(b)}$ The data was processed in modified HRSDI (annularly)  (see, Section~\ref{sec:cal-issues} for details).}

%\end{tablenotes}
\end{table*}

%%%%%%%%%%%%%%%%%%%%%%%%%%%%%%%%%%%%%%%%%%%%%%%%

%\subsection{Solving the small sample statistic problem}
\subsection{Measuring the noise at small angular separations}

At small angular separations only a few statistically independent apertures can be made. This makes it difficult to get an accurate estimate of the noise. We can solve the problem of  small sample statistics  by increasing the size of our sample. In other words, the problem will be solved if we can make enough noise measurements. However, this is impossible for high-contrast imaging done with classical (narrow- or broadband) imagers. On the other hand, medium-resolution IFS provides alternative solutions by 
(i) measuring the noise in the spectral direction or
(ii) measuring the noise in the spatial direction at multiple wavelengths. It should be noted that methods (i) and (ii) reach the same answer if the PSF is circular symmetric because MUSE can reach the photon noise limit at small separations.

At small separations, the noise in Equation~\eqref{equ_noise_component} can be simplified as photon noise dominant $\sigma_{\lambda}$ = (c$_\lambda \cdot F^{\alpha}_\lambda$). Hence, we can compute the noise at \Ha~based on the noise we measured at multiple wavelengths at the same separation as
\begin{equation}
\label{equ_noise_Ha}
\sigma_{H\alpha} = \sigma_{ \lambda} (\frac{c_{H\alpha} \cdot F^{\alpha_2}_{H\alpha}}{c_{\lambda}\cdot F^{\alpha_1}_{\lambda}}),
\end{equation}
where the coefficient $c$ and exponent $\alpha$ at corresponding wavelength are obtained from the result of two noise components fitting. It is worth  noting that Equation~\eqref{equ_noise_Ha} can be used to estimate the noise at any wavelength, not just at \Ha.

There are approximately 6, 12, and 18 resolution elements at 1 $\lambda$/D, 2 $\lambda$/D, and 3 $\lambda$/D, respectively. In practice, at least 20 independent measurements of the noise are recommended to avoid the penalty induced by small sample statistics \citep{Mawet2014}. It is impossible to make 20 noise measurements at \textless 3 $\lambda$/D on the \Ha~image. However, even with only one measurement at one wavelength, we can easily find more than 20 wavelengths to perform the noise measurement and derive the correspond noise at \Ha~using Equation~\eqref{equ_noise_Ha}. 

At small separations (\textless~3 $\lambda$/D), using Equation~\eqref{equ_noise_Ha} can provide direct measurement of the noise to avoid the small sample statistics problem. At larger separations (\textgreater~3 $\lambda$/D), both Equation~\eqref{equ_noise_Ha} and the classical approach of measuring the noise in the same image as the detection can give accurate estimate of the noise, assuming the photon noise dominant (see Fig.~\ref{contrast_curve_new}). However, the presence of extended structures (such as stellar jets) may reduce the empty regions where we can measure the noise. In this case, Equation~\eqref{equ_noise_Ha} will become useful since the atomic jets are only present at a few spectral channels. Throughout the paper Equation~\eqref{equ_noise_Ha} was used to estimate the noise when building the contrast curve on images with photon-noise dominant residuals.

%%%%%%%%%%%%%%%%%%%%
%%%%%%%%%%

\subsection{Contrast curve}

%%%%%%%%%%%%%%%%%%%%
\begin{figure}
\centering
\includegraphics[width=0.48\textwidth]{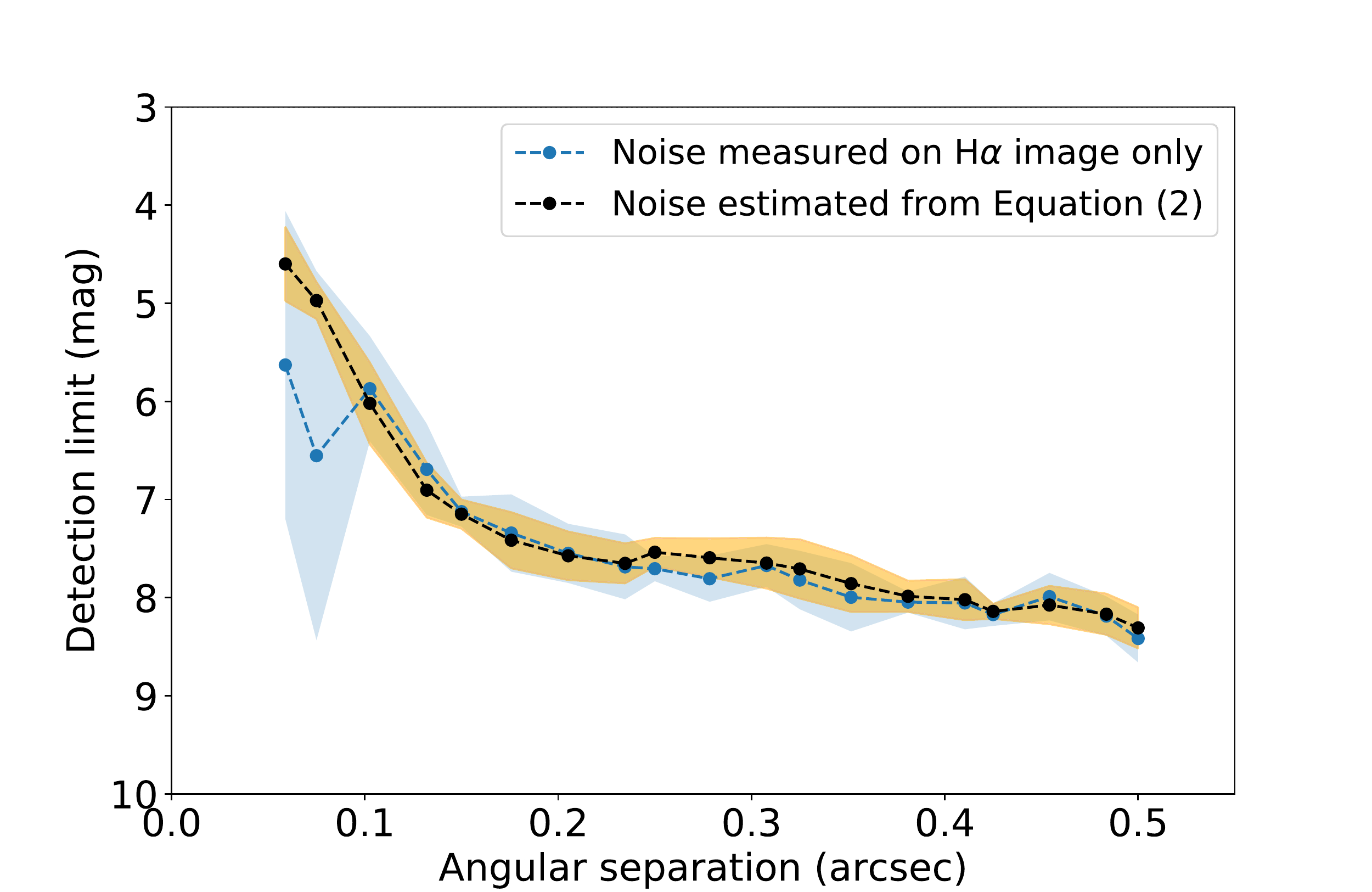}
%\vspace{-8mm}
\caption{Contrast curves for J1850-3147. The colored shaded regions around each curve represent the standard deviation of the contrasts obtained at eight PAs.}
\label{contrast_curve_new}
\end{figure}
%%%%%%%%%%

%Figure~2 shows the 5$\sigma$ contrast curve of CD-31 16041. 
Figure~\ref{contrast_curve_new} shows two H$\alpha$ contrast curves for J1850-3147 at 5$\sigma$ confidence level. The black curve is created based on the noise estimated from Equation~\eqref{equ_noise_Ha}. As a comparison, the blue curve is obtained with noise measured on \Ha~image only, which is affected by small sample statistics at \textless~0.1\arcsec. The colored shaded region indicates the uncertainty of the contrast curve. 

\begin{figure}
\centering
\includegraphics[width=0.45\textwidth]{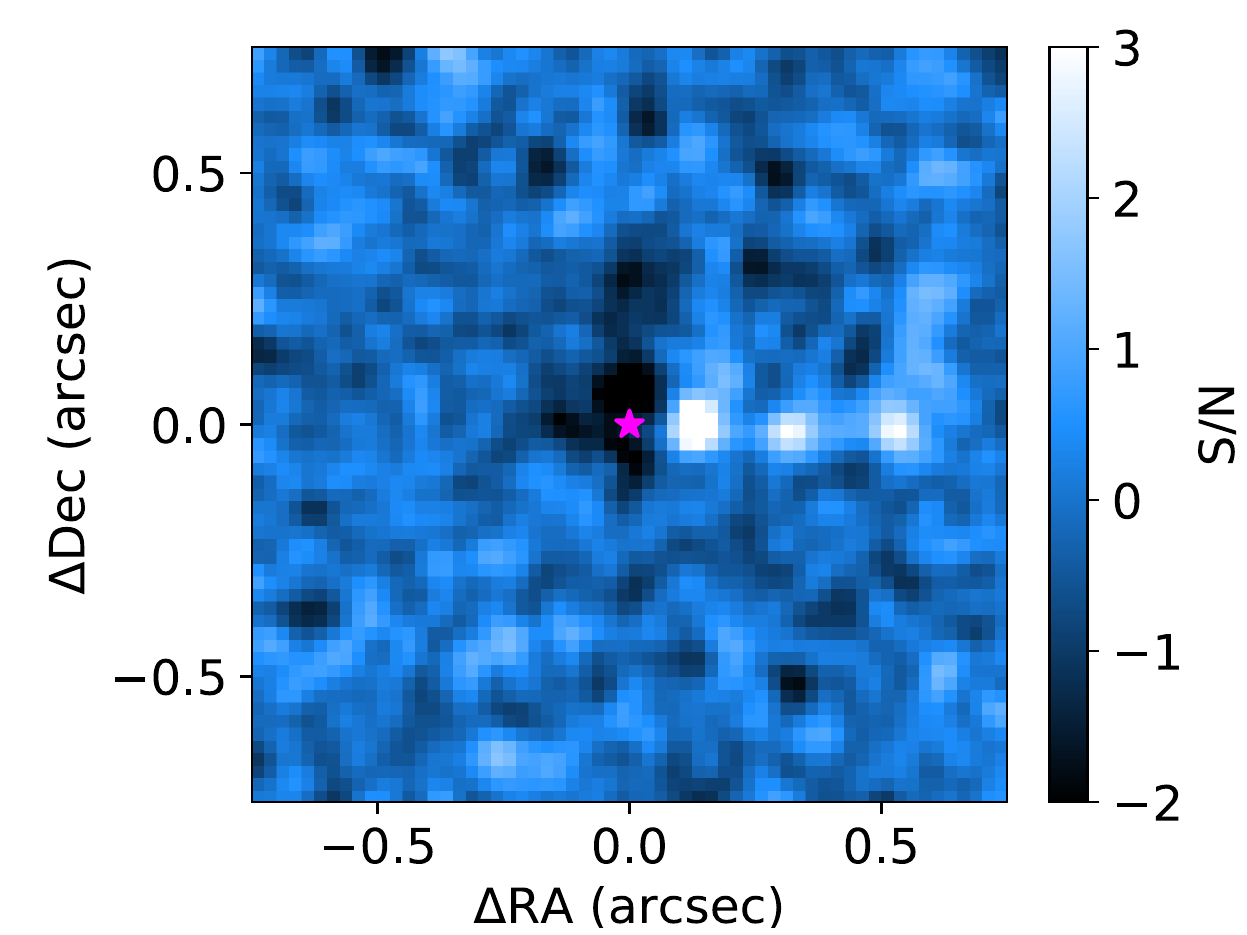}
%\vspace{-8mm}
\caption{S/N map of J1850-3147 at \Ha~after the process of HRSDI is shown. For visual inspection of the contrast curve derived from Equation~\eqref{equ_noise_Ha}, three 5$\sigma$ fake planets were injected at  distances of 0.1\arcsec, 0.3\arcsec, and 0.5\arcsec. All three fake planets are clearly visible. For display purposes, the image was smoothed with a Gaussian kernel of 2 pixels in radius. The magenta star indicates the star center.    }
\label{visual_inspection}
\end{figure}

No planet is found in J1850-3147. The 3 min MUSE observation is sensitive to objects 7 mag fainter than the host star at the same wavelength at a separation of 150 mas. Overall, Equation~\eqref{equ_noise_Ha} provides more robust noise estimations as it has consistent results with direct measurement on \Ha~image, but with much smaller scatter at small separations (\textless~0.15\arcsec). The separation of 0.15\arcsec equals two resolution elements of MUSE\footnote{The resolution element of MUSE is defined by FWHM, not $\lambda$/D, because currently the AO system installed on VLT cannot reach the diffraction limit at optical wavelengths.}. For the visual inspection of Equation~\eqref{equ_noise_Ha}, we injected the fake planet at three different locations (r = 0.1\arcsec, 0.3\arcsec, and 0.5\arcsec) at 5$\sigma$ confidence level. All three injected planets in Fig.~\ref{visual_inspection} are clearly visible, as suggested by the contrast curve. 

%%%%%%%%%%%%%%%%%%%%
%%%%%%%%%%

\subsection{Detection limits in \texorpdfstring{H$\alpha$} line flux}

%%%%%%%%%%%%%%%%%%%%
\begin{figure}
\centering
\includegraphics[width=0.5\textwidth]{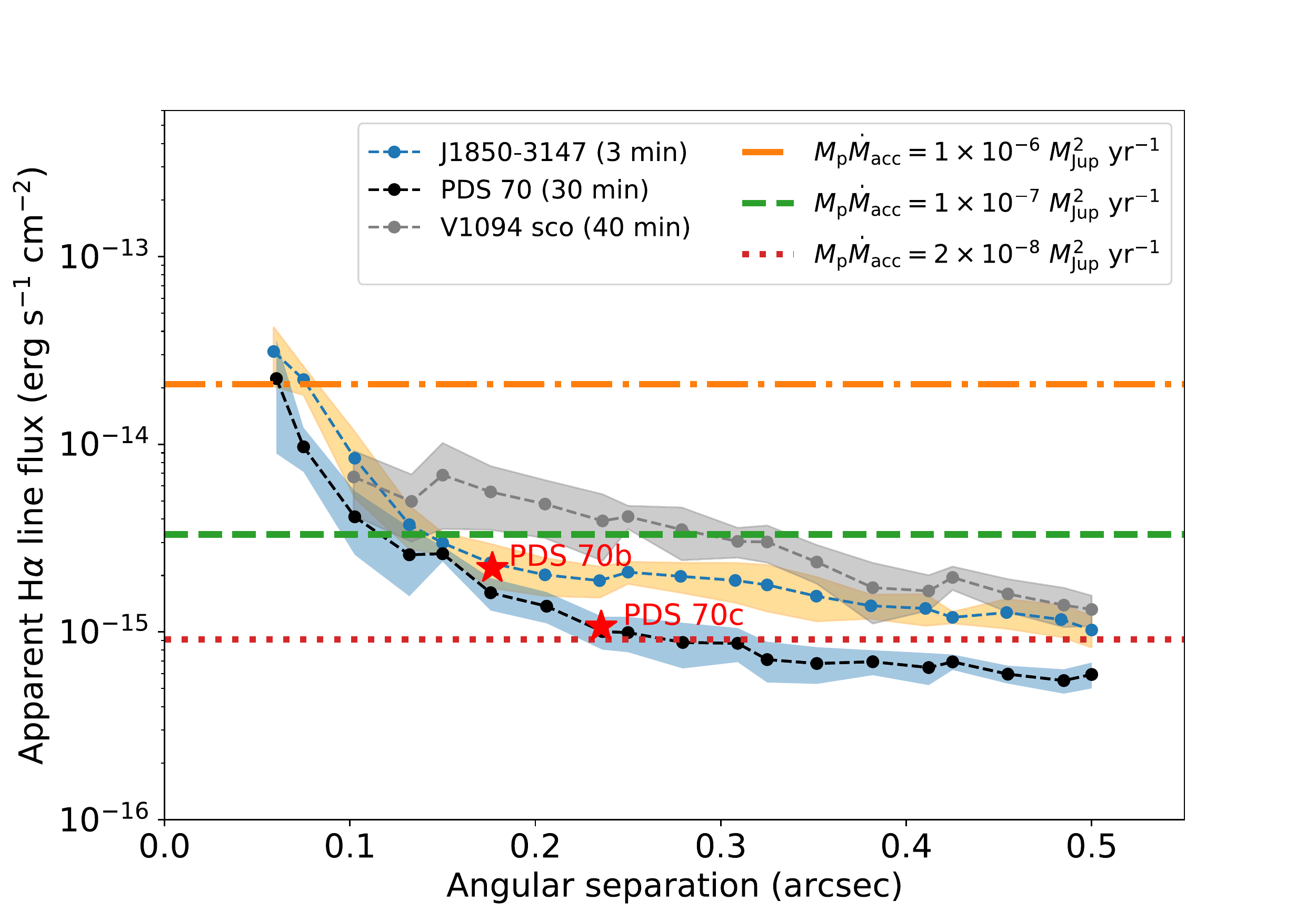}
%\vspace{-8mm}
\caption{Detection limits in apparent \Ha~line flux at 5$\sigma$ confidence levels. The colored shaded regions around each curve represent the standard deviation of the contrasts obtained at eight PAs. PDS 70 b and c from \cite{Haffert2019} are shown as red stars. The three horizontal lines show predicted \Ha~emission estimated from \cite{Gullbring1998, Rigliaco2012}. The detailed estimation can be found in Sect.~\ref{sec:predict_flux} }
\label{line_flux_Macc}
\end{figure}
%%%%%%%%%%

%The sensitivity of apparent line flux can be used to make comparison between various instruments and targets. 

\begin{figure*}
   \centering
    \includegraphics[width=0.95\textwidth]{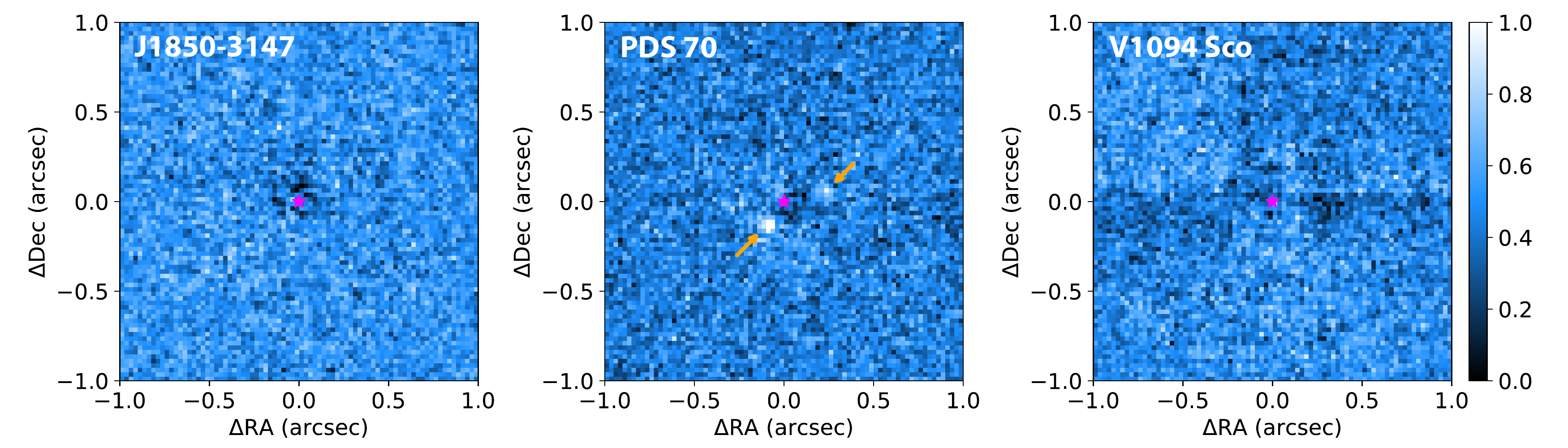}
      \caption{\Ha~detection maps after the removal of starlight. PDS b and c are indicated by the yellow arrows. The magenta star gives the position of the star.
      }
         \label{final_images}
\end{figure*}

To better illustrate the performance of MUSE, we present the detection limit in apparent flux at 5$\sigma$ confidence level (see Fig.~\ref{line_flux_Macc}). The colored shaded region indicates the uncertainty of the detection limit. %The detection limit in apparent line flux is directly linked to the required accretion activities for the different mass of planets. 
Due to the instrumental issues in MUSE (described in Sect.~\ref{sec:cal-issues}), we could only apply HRSDI in three out of five targets to perform high-contrast imaging. The \Ha~detection maps\footnote{The detection map is essentially the S/N map with the scale linearly shifted to an arbitrary unit from 0 to 1.} for these targets are shown in Fig.~\ref{final_images}.

In general, a 30 min MUSE observation can detect apparent \Ha~line flux down to 10$^{-14}$ and 10$^{-15}$ erg~s$^{-1}$~cm$^{-2}$ at 0.075\arcsec~and 0.25\arcsec, respectively. Since MUSE can be photon noise dominated (\textless~1\arcsec),  we expect the detection limit to go down with the square root of integration time, which is supported by datasets with different integration times (see Fig.~\ref{line_flux_Macc}). Below we introduce the detection limit of each target.

~\par
\noindent {\sl J1850-3147}

~\par
The detection limit of J1850-3147 is indicated by the blue curve with yellow shaded regions. Even with 3 min integration time, MUSE can already reach the \Ha~line flux of $3 \times 10^{-15}$ erg~s$^{-1}$~cm$^{-2}$ at 150 mas and $3 \times 10^{-14}$ erg~s$^{-1}$~cm$^{-2}$ at 50 mas with 5$\sigma$ confidence level. %PDS~70~b was detected by MUSE at 11$\sigma$ with 30 min integration time, which has \Ha~line flux of $3.9 \times 10^{-16}$ erg~s$^{-1}$~cm$^{-2}$ \citep{Haffert2019}. 
In the case of J1850-3147, we can detect PDS~70~b (if present) at around 5$\sigma$ confidence level  with 3 min observations (see Fig.~\ref{line_flux_Macc}).

~\par
\noindent {\sl PDS 70}

~\par
The detection limit of PDS 70 data used in \cite{Haffert2019} is represented by the black curve in Fig.~\ref{line_flux_Macc}. The integration time of this data is 30 min. The regions where two planets were located were excluded during the analysis of fake planet injection. Except for PDS~70~b and c, we do not find other planets with 5$\sigma$ upper limit of \Ha~line flux of $2.5 \times 10^{-15}$ erg~s$^{-1}$~cm$^{-2}$ at 150 mas and $2 \times 10^{-14}$ erg~s$^{-1}$~cm$^{-2}$ at 50 mas. 
Overall, the detection limit of PDS~70 is two to three times deeper than that of J1850-3147, which is consistent with the expected S/N gain of $\sqrt{(30~ \rm min)/(3~ \rm min)}$. 

~\par
\noindent {\sl V1094~Sco}

~\par

V1094~Sco was significantly affected by instrumental issues in MUSE (see Sect.~\ref{sec:cal-issues}). However, the V1094 Sco data can be processed with modified HRSDI by building the normalized reference spectra annularly (see Sect.~\ref{sec:modified_HRSDI}). 

The detection limit in apparent \Ha~line flux for V1094~Sco is shown in Fig.~\ref{line_flux_Macc}. At the location of the gap 1 ($\sim$0.5\arcsec) identified by \cite{vanTerwisga2018}, we do not find any planets. The MUSE observation yields a 5$\sigma$ upper limit in \Ha~line flux of about $1.3 \times 10^{-15}$ erg~s$^{-1}$~cm$^{-2}$. Due to the poor seeing conditions during the observations, the FWHM of V1094~Sco is about 150 mas, which is roughly two times larger than that of J1850-3147 and PDS~70. Therefore, the detection limit curve is flattened compared with the other two targets. Although we expect the detection limit to go down with the square root of integration time, the instrumental issues and the poor atmospheric conditions performance will also limit the performance of MUSE. As a result, the detection limits for V1094~Sco and J1850-3147 are similar.

\subsection{Comparison with SPHERE/ZIMPOL}

\begin{figure}
   \centering
    \includegraphics[width=0.48\textwidth]{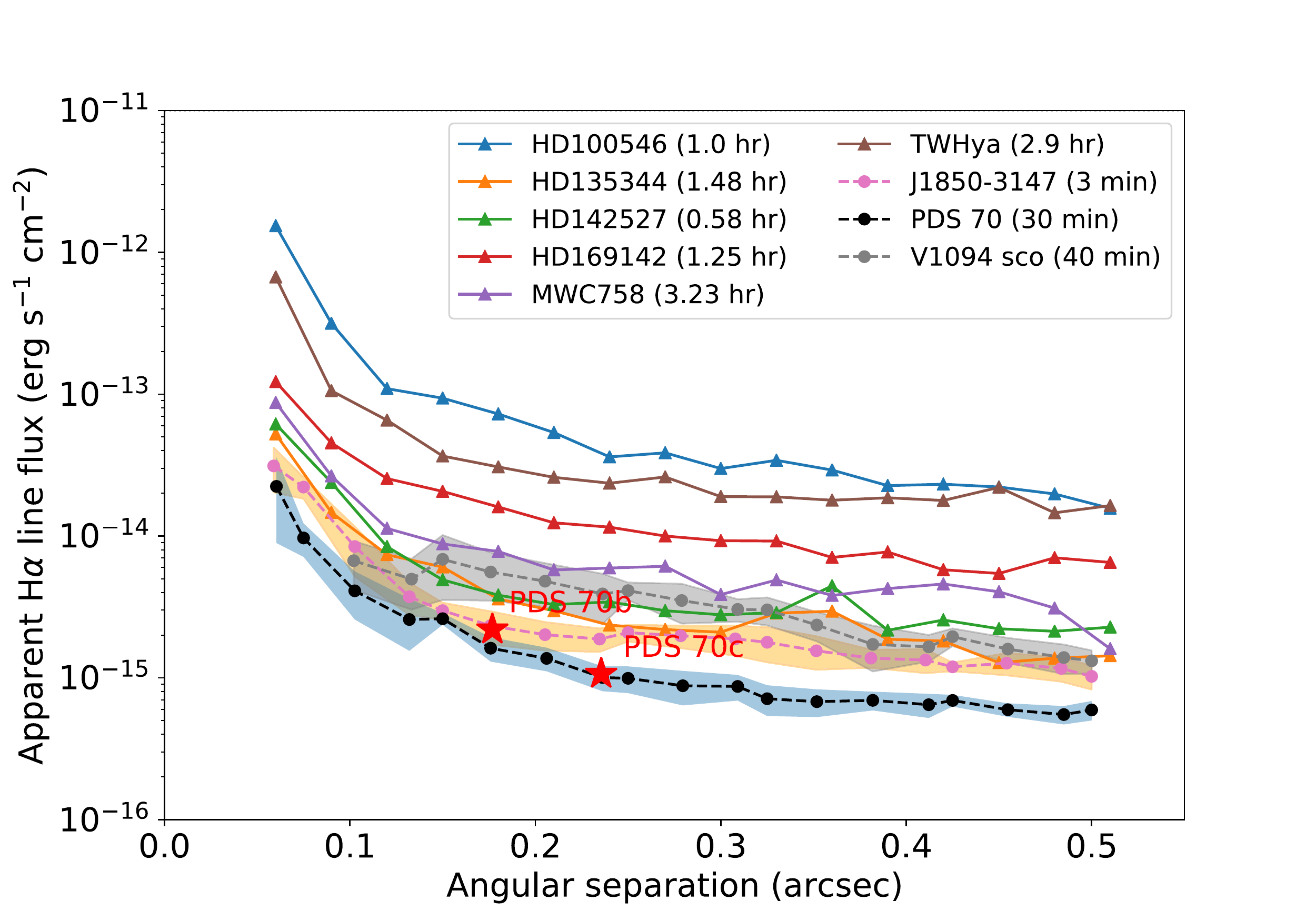}
      \caption{Detection limits (5$\sigma$) in apparent flux obtained with MUSE (dashed lines with colored shaded regions), together with the detection limits of ZIMPOL \citep[solid lines with triangle markers;][]{Cugno2019}. The total integration time is listed after the target name. PDS 70 b and c from \cite{Haffert2019} is also shown.
      }
         \label{compare_app_flux_SPHERE}
\end{figure}

In Fig.~\ref{compare_app_flux_SPHERE}, we present the comparison between MUSE and SPHERE/ZIMPOL in terms of the detection limit in apparent \Ha~line flux. The ZIMPOL results are taken from \cite{Cugno2019}. In general, the MUSE detection limit in apparent \Ha~line flux is a factor of 5 times deeper than the detection limit of ZIMPOL due to the gain in resolving power. More importantly, these detection limits were achieved with 3 to 40 min of MUSE observations. Meanwhile, the total integration times used in \cite{Cugno2019} were ranging from 40 min to 3 hr, as labeled in Fig.~\ref{compare_app_flux_SPHERE}. In other words, MUSE can reach higher S/N than ZIMPOL in a shorter integration time. Moreover, without the laser-guided AO system, it is very difficult for ZIMPOL to observe faint targets such as PDS 70 and V1094 Sco. We discuss the potential reasons that cause this improvement in Sect.~\ref{sec:reasons}.

%%%%%%%%%%%%%%%%%%%%%%%%%%%%%%%%%%%%%%%%%%%%%%%%%%

\subsection{Mapping stellar jets with MUSE}
\label{sec:mping_jets}

\begin{figure}
   \centering
    \includegraphics[width=0.48\textwidth]{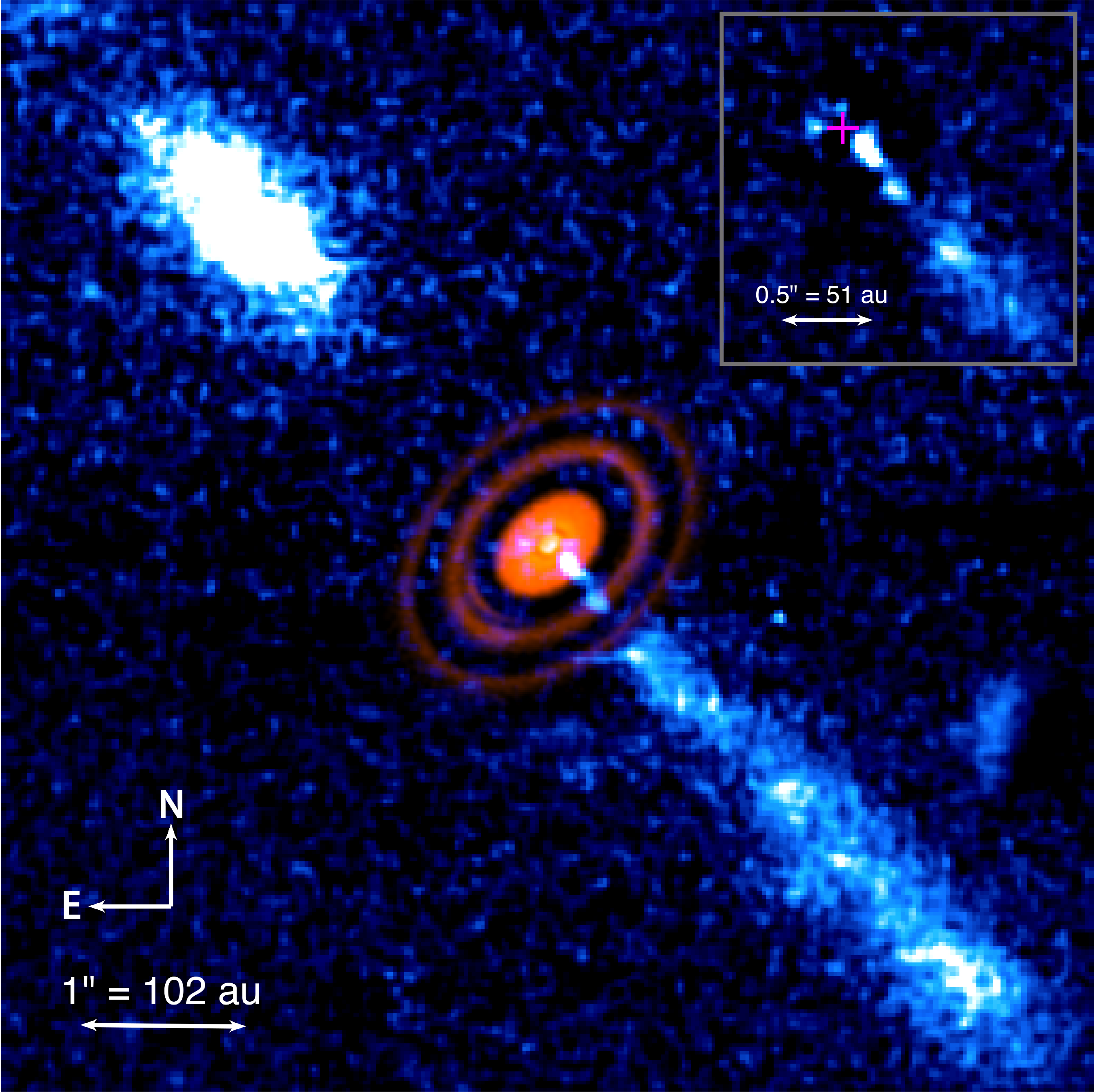}
      \caption{Composite image of HD 163296 shows two jets detected by MUSE, superimposed to the disk detected by ALMA at 1.2 mm \citep{DSHARP_I_Andrews2018, DSHARP_IX_Isella2018}. The NE jet and SW jet are traced by [SII] 673nm and \Ha~lines, respectively, and are smoothed with a Gaussian kernel of 2 pixels in radius. The arc-like structure near the SW jet is the blob ghost caused by the instrument (see Appendix~\ref{sec:blob_ghost}). The inset box of \Ha~image shows the inner most region. The magenta cross marks the stellar position. 
      }
         \label{jets_cnt}
\end{figure}

Figure~\ref{jets_cnt} presents a composite image of HD~163296 showing two jets from this work and the disk detected by ALMA at 1.2 mm \citep{DSHARP_I_Andrews2018, DSHARP_IX_Isella2018}. %After removing the stellar emission, 
The northeast (NE) jet is redshifted and contains one elongated knot, traced by the [S II] $\lambda$673~nm line.  The southwest (SW) jet is blueshifted and has a continuous stream with a length of $\sim$450~au, traced by the \Ha~line. Different atomic lines were detected in each jet and show consistent morphology. The asymmetry of the two jets observed by MUSE is similar to the VLT/Xshooter observation at separations larger than 2\arcsec, which show knots and a relatively continuous stream on the redshifted and blueshifted jet, respectively \citep{Ellerbroek2014}.

The jets were detected in MUSE observations with exposure setting of 8~s and 250~s and with multiple position angles (see Table~\ref{tab:obs_log}). To make the composite image in Fig.~\ref{jets_cnt}, the inner saturated region in the long exposure dataset was masked and replaced with short exposure dataset that only has minor saturation in the most central few spaxels. Without the coronagraph, MUSE is able to probe the jets down to 1 FWHM of $\sim$75 mas, corresponding to 8 au at the target distance of 101.5 pc. Due to the minor saturation in the short exposure (8~s) dataset, anything within 100 mas should be treated cautiously. 

%\subsection{2D morphology in South-west blueshifted jet}

The inset  in Fig.~\ref{jets_cnt} shows the jet image zoomed in at the center. The resolution of MUSE is about 8 au at \Ha. %The spatially resolved detection of the blueshifted \Ha~jet suggests the launching site is less than 100 mas or 10 au. %The jet width measured from \Ha~is approximately 12 au within 60 au (0.6\arcsec) from the star and becomes wider when it propagates outwards (see Fig.~\ref{jets_cnt}). 
There are three bright spots on top of the faint stream, indicating that there is a potential dynamical eruption history. Detailed analysis of the stellar jets in HD~163296 will be presented in future work.

%%%%%%%%%%%%%%%%%%%%%%%%%%%%%%%%%%%%%%%%%%%%%%%%
\begin{table}[th!]
\caption{Targets properties}             % title of Table
\label{tab:target_flux}      % is used to refer this table in the text
\centering                          % used for centering table
\begin{tabular}{c c c c }        % centered columns (4 columns)
\hline\hline                 % inserts double horizontal lines
Name &  H$_{\alpha}$ & $F_{\rm H\alpha}$ & ratio \\    % table heading 
  & (mag) & (10$^{-13}$ erg/s/cm$^{2}$) & \\
\hline                        % inserts single horizontal line
PDS 70   & 16.71 & 7.5 $\pm$ 0.4 &  1.60   \\      % inserting body of the table
J1850-3147   & 16.23 & 9.1 $\pm$ 0.8 & 1.42   \\
V1094 Sco   & 18.08 & 6.3 $\pm$ 0.1 & 2.47   \\
HD 100546   & 11.51 & 4312.5 $\pm$ 35.3 & 5.63    \\
HD 163296   & 12.13 & 1799.0 $\pm$ 43.0 & 3.28    \\

\hline%inserts single line
\end{tabular}
%\begin{tablenotes}
\tablefoot{Columns 2 and 3 give the apparent stellar \Ha~flux observed by MUSE. We adopt the narrowband filter (N{\_}Ha) used in ZIMPOL to measure the stellar brightness at \Ha, listed in column 2. We convert the flux unit to AB magnitude system using, ${\displaystyle m_{\text{AB}}\approx -2.5\log _{10}\left({f_{\nu }}/{(3631{\text{ Jy}})}\right).}$ The $F_{\rm H\alpha}$ is line flux obtained from single Gaussian fitting. Column 4 lists the ratio of peak line flux to continuum flux. All the measurements are observed results without the correction for possible dust extinction from the disk.
}
 %(erg/s/cm$^{2}$)
% $1.75 \pm 0.46$ $\times$ 10$^{11}$
%\item {a.} From Planck measurements \cite{planck15}.
%\end{tablenotes}
\end{table}  
%%%%%%%%%%%%%%%%%%%%%%%%%%%%%%%%%%%%%%%%%%%%%%%% 

%%%%%%%%%%%%%%%%%%%%%%%%%%%%%%%%%%%%%%%%%%%%%%%%%%

%%%%%%%%%%%%%%%%%%%%%%%%%%%%%%%%%

\section{Discussion}
\label{sec:discussion}

\subsection{Reasons for the improvement}
\label{sec:reasons}

There are multiple reasons that account for the better performance of MUSE compared to SPHERE/ZIMPOL. Even though they are different instruments   requiring different post-processing techniques, we can still reach a quantitative reason and some qualitative expansions.

The primary reason that we can quantify is that MUSE has a spectral resolution of 1.25~\AA, while ZIMPOL used much wider narrow- and broadband filters. For photon-noise dominant residuals, the S/N scales with the square root of the spectral resolving power and the telescope diameter. In this work we combine three spectral channels (total 3.75 \AA) to match the line width of injected fake planets. \cite{Cugno2019} used a broadband \Ha~filter (B{$\_$}Ha) with the width of 5.5~nm and a continuum filter (Cnt{$\_$}Ha) centered at 644.9~nm\footnote{The detail of central wavelengths and filter widths can be found in Table~6 of \cite{Schmid2018}}. With equal mirror diameter, the S/N of MUSE compared to ZIMPOL scales approximately as $\sqrt{(55 \text{\AA})/(3.75 \text{\AA})}$~$\sim$~4 for using the B{$\_$}Ha filter, assuming ZIMPOL is also photon noise dominant. %MUSE is photon noise dominant at small separation to sky background limited at large separation.  
In principle, using a narrowband filter could improve the S/N of detecting \Ha~line emission as it includes less noise from continuum emission. However, the assumption of photon noise dominant is not always true for ZIMPOL. At the separations \textgreater 0.3\arcsec, the data from the narrowband filter (N{$\_$}Ha) in ZIMPOL can be read-noise limited. As suggested in \cite{Cugno2019}, further studies are needed to determine the relation between longer integration time and detection limit of ZIMPOL.

The secondary reason is the lack of  detailed spectral information of the primary star. It can lead to oversubtraction when removing continuum emission using the Cnt{\_}Ha filter. For A/B type  stellar spectra, the spectral indices\footnote{The spectral index $\alpha$ is defined as $F_{v} \propto v^{ \alpha}$} are usually greater than 1 around \Ha. As a result, the oversubtracted flux has to be compensated by line emission. Three out of four targets in \cite{Cugno2019} that have high limits of line flux are A/B type stars, and the integration times for these targets are relatively long, ranging from 1 to 3 hr.

% using flux-scaling stellar PSF, which contained both continuum and line emission.

Another reason is how the contrast analysis was performed. \cite{Cugno2019} created the fake planet with the stellar image obtained by filters. Although using the stellar PSF as the planet PSF is a common approach, using filters can only scale the flux of the stellar PSF that contains both continuum and line emission. As for MUSE, we used a simple Gaussian line profile as the injected planet spectrum, which only has emission at \Ha, as mentioned in Sect.~\ref{sec:planet_inj}. In this way we can only inject line emission and determine the detection limit in line flux more accurately.

% different throughput and AO system under different seeing conditions

SPHERE is designed to image bright stars (e.g., m$_{R}$ \textless~10~mag; \cite{Beuzit2019}). Its Strehl ratio drops quickly from 60\% at m$_{R}$ \textless~9~mag to 10\% at m$_{R}$ = 12~mag. Without the laser-guided AO system, it is very difficult for SPHERE to observe faint stars and still achieve a high Strehl ratio. A low Strehl ratio could explain the high detection limit of TW Hya, which is the faintest target (m$_R$ = 10.43) with the worst seeing ($\sim$1.3\arcsec) conditions in \cite{Cugno2019}.

\subsection{Constraining the planet mass and the mass accretion rate}
\label{sec:predict_flux}

\Ha~emission originates from accretion activity. The line flux and the line width can be used to estimate the mass accretion rate. Currently there are two methods commonly used to estimate the mass accretion rate of young stars, which we use to estimate the mass accretion for planets.% Below we introduce those methods in some more detail. 
\par
~\par

\noindent \textbf{Line width:} The mass accretion rate can be computed from the line width as 
\begin{equation}
\label{equ_Macc_line_width}
{\rm log}(\dot{M}_{\rm acc}) = -12.89(\pm0.3) + 9.7(\pm0.7) \times 10^{-3}~{\rm H}\alpha(10\%),
\end{equation}
where H$\alpha$(10\%) is the 10\%~width of the \Ha~line in km~s$^{-1}$ and the unit of $\dot{M}_{\rm acc}$ is M$_{\odot}$~yr$^{-1}$ \citep{Natta2004}.

~\par
\noindent \textbf{Line luminosity:} The mass accretion rate can also be converted from the accretion luminosity, which has the empirical relation \citep{Herczeg2008, Fang2009, Rigliaco2012}  
\begin{equation}
\label{equ_accretion_luminosity}
{\rm log}(L_{\rm acc}) = b + a{\rm log}(L_{{\rm H}\alpha}), 
\end{equation}
where  $L_{{\rm H}\alpha}$ is the line luminosity. We adopted $a$ and $b$ as 1.25 and 2.27 \citep{Fang2009}, respectively. The mass accretion rate is
\begin{equation}
\label{equ_Macc_line_luminosity}
\dot{M}_{\rm acc} = (1-\frac{R_{\rm *}}{R_{\rm in}})^{-1} \frac{L_{\rm acc} R_{\rm *}}{G M_{\rm *}}, 
\end{equation}
where  $R_{\rm *}$ and $M_{\rm *}$ are the planet radius and its mass, and $R_{\rm in}$ is the inner radius of accretion disk \citep{Gullbring1998}. 

Both methods were explored by fitting the data of low-mass stars or brown dwarfs \citep{Natta2004, Fang2009}, but not planets. %However, those were the best knowledge we had so far to predict the mass accretion rate of the planet. 
With the actual data we obtained with MUSE, we can start to make constraints on those accretion models on planetary scales \citep[e.g.,][]{Thanathibodee2019}. Throughout this paper we follow \cite{Gullbring1998} to estimate the line luminosity with given planet mass and $\dot{M}_{\rm acc}$.

In Fig.~\ref{line_flux_Macc} we present the predicted apparent \Ha~flux (horizontal lines) with planet mass of 1~$M_{\rm Jup}$ and multiple mass accretion rates $\dot{M}_{\rm acc}$, assuming the host star at 140 pc (the distance to nearby star forming regions) without the disk extinction. %\footnote{Without any prior information, A$_{v}$ of 3 mag is a relatively conservative value.}. 
In general, MUSE can detect Jupiter-mass planets with a very low mass-accretion rate (e.g., $2 \times 10^{-8} M_{\rm Jup}$~yr$^{-1}$) in a 30 min integration time (see Fig.~\ref{line_flux_Macc}), which is impossible for any other current instrument. With such sensitivity, non-detections can also provide meaningful upper limits that are $\sim$5 times deeper than current results, which is important for the future study of accretion variability and provides a constraint on the planet mass.

Based on MUSE measurements of \Ha~line intensities and \Ha~line widths, we can decouple the planet mass and the mass accretion rate estimated by Equations~\eqref{equ_Macc_line_width} and~\eqref{equ_Macc_line_luminosity}, with minor assumptions on the planet radius and the inner radius of circumplanetary disks (CPD). %The mass accretion rate $\dot{M}$ can be estimated from \Ha~line width as in Equation~\eqref{equ_Macc_line_width}. Based on  Equation~\eqref{equ_Macc_line_luminosity} the product of planet mass and its accretion rate $M_{\rm p}\dot{M}$ can be estimated based on \Ha~line flux. 
It is very powerful that a single MUSE observation can provide estimations of the planet mass and the mass accretion rate, although the uncertainty of the mass accretion rate could be as large as one order of magnitude due to the uncertainty of accretion models and limited spectral resolution of MUSE \citep{Thanathibodee2019}. Nevertheless, recent radiation-hydrodynamic models made by \cite{Aoyama2019} demonstrated that combining \Ha~line widths and line intensities could narrow down the ranges of the planet mass and the accretion rate significantly. Furthermore, adding information from other lines, such as H$\beta$, can put an even better constraint on the accretion properties \citep{Hashimoto2020}.

%\subsubsection{Detection limit of H$\beta$}

\subsection{Observation strategies for target selection and detecting planetary \texorpdfstring{H$\beta$} line}

Young stars (<10 Myr) with weak \Ha~line emission and faint {\sl R} band brightnesses are suitable targets for future MUSE survey of protoplanets. This is  due to two technical reasons. On the one hand, MUSE has instrumental issues that are strongly affected by the \Ha~line-to-continuum ratio, as mentioned in Sect.~\ref{sec:cal-issues}. Hence, choosing weak \Ha~emitters can avoid being limited by systematic noise caused by instrumental issues. Nevertheless, targets with a medium \Ha~line-to-continuum ratio can still be processed with modified HRSDI. On the other hand, MUSE is designed to take long exposures. Although the overhead between each exposure is reduced to $\sim$15~s, it is still not very efficient for observing bright stars as they require a very short integration time (a few seconds) to avoid  saturation. %As a result, inefficient science observation. 

%%%%%%%%%%%%%%%%%%%%
%%%%%%%%%%
% AO performance PSF 
In addition to the \Ha~line, another strong accretion signature is H$\beta$. Detecting the H$\beta$ line   can help to understand the accretion mechanism as it predicts the presence of the H$\beta$ emission line, and it can also provide a direct measurement of the extinction of CPD and circumstellar disk.

In general, in order to detect the H$\beta$ we suggest a total integration time that is at least a factor of 36 longer than that  needed for \Ha.\   Usually, the line flux of H$\beta$ is 3 times lower than that of \Ha. To compensate the slightly worse Strehl ratio at the shorter wavelength (i.e., H$\beta$), a factor of 2 is needed to achieve the same detection limit as in \Ha. Therefore, to observe the H$\beta$ line, we need to go 6 times deeper than observing \Ha~line, corresponding to 36 times longer integration time under the assumption of photon noise dominant residuals.

\section{Conclusions}
\label{sec:conclusions}

In this paper we present the performance of MUSE as a high-contrast imager in detecting point-like planets and extended stellar jets by analyzing MUSE observations of five stars with various brightnesses and spectral types. The results are summarized below:

\begin{enumerate}

\item We found at least three noise components in the MUSE data: photon noise, sky background noise, and systematic noise. For the weak \Ha~emitter, systematic noise can be ignored. In this  case, MUSE can reach the photon noise limit at small separations (\textless 0.5\arcsec) and be sky background limited at the outer regions (\textgreater 1.5\arcsec), after the process of HRSDI.  

\item MUSE can avoid the small sample statistics problem by measuring the noise in spatial direction at multiple wavelengths. The noise derived from Equation~\eqref{equ_noise_Ha} can provide robust noise estimations at the small sample regime and provide consistent noise estimations outside the small sample regime.  

\item In general, a 30 min MUSE observation can detect 5$\sigma$ planets with apparent \Ha~line flux down to 10$^{-14}$ and 10$^{-15}$ erg~s$^{-1}$~cm$^{-2}$ at 0.075\arcsec~and 0.25\arcsec, respectively. Despite the analyses based on different targets with different integration time, MUSE detection limits in apparent flux are about 5 times deeper than that of ZIMPOL.

\item Except for PDS~70~b and~c, we did not detect any clear accretion signature in PDS~70, J1850-3147, and V1094~Sco down to 0.1\arcsec.

\item We detected two asymmetric atomic jets in HD~163296 with a very high spatial resolution (down to 8 au) at optical wavelengths. %We identified five atomic lines [O I] $\lambda$630~nm, [N II] $\lambda$658~nm, \Ha, [S II] $\lambda$671~nm, and [S II] $\lambda$673~nm in the jets. The radial velocities of redshifted and blueshifted jets are around 120-140 km~s$^{-1}$ and -270 km~s$^{-1}$, respectively.

\end{enumerate}

MUSE has instrumental issues, and  the stellar \Ha~line-to-continuum ratios vary significantly across the fields, which can potentially limit its performances. Objects with higher \Ha~line-to-continuum ratio are affected more severely by variations. A proposed modified HRSDI that adjusts the way of building the normalized reference spectra can minimize the instrumental issues. For future searches for protoplanets with MUSE, we suggest young stars with a weak \Ha~line emission and a faint $R$-band magnitude (m$_R$ > 10 mag) to be the suitable targets to avoid instrumental effects altogether.

\begin{acknowledgements}
We thank Roland Bacon and Peter Weilbacher for insightful discussions about ghosts in MUSE. This work is based on observations collected at the European Organisation for
Astronomical Research in the Southern Hemisphere under ESO programmes
60.A-9100(K), 60.A-9482(A), and 0103.C-0399(A). This research made use of Astropy,\footnote{http://www.astropy.org} a community-developed core Python package for Astronomy \citep{astropy:2013, astropy:2018}.

Support for this work was provided by NASA through the NASA Hubble
Fellowship grant \#HST-HF2-51436.001-A awarded by the Space Telescope
Science Institute, which is operated by the Association of Universities for Research in Astronomy, Incorporated, under NASA contract NAS5-26555. I.S. acknowledges funding from the European Research Council (ERC) under the European Union's Horizon 2020 research and innovation program under grant agreement No 694513. JB acknowledges support by FCT/MCTES through national funds by the grant UID/FIS/04434/2019, UIDB/04434/2020 and UIDP/04434/2020 and through the Investigador FCT Contract No. IF/01654/2014/CP1215/CT0003.

\end{acknowledgements}

%-------------------------------------------------------------------
\bibliographystyle{aa}
\bibliography{ms}

\begin{thebibliography}{105}
\expandafter\ifx\csname natexlab\endcsname\relax\def\natexlab#1{#1}\fi

\bibitem[{{Alcal{\'a}} {et~al.}(2017){Alcal{\'a}}, {Manara}, {Natta}, {Frasca},
  {Testi}, {Nisini}, {Stelzer}, {Williams}, {Antoniucci}, {Biazzo}, {Covino},
  {Esposito}, {Getman}, \& {Rigliaco}}]{2017A&A...600A..20A}
{Alcal{\'a}}, J.~M., {Manara}, C.~F., {Natta}, A., {et~al.} 2017, \aap, 600,
  A20

\bibitem[{{ALMA Partnership} {et~al.}(2015){ALMA Partnership}, {Brogan},
  {P{\'e}rez}, {Hunter}, {Dent}, {Hales}, {Hills}, {Corder}, {Fomalont},
  {Vlahakis}, {Asaki}, {Barkats}, {Hirota}, {Hodge}, {Impellizzeri}, {Kneissl},
  {Liuzzo}, {Lucas}, {Marcelino}, {Matsushita}, {Nakanishi}, {Phillips},
  {Richards}, {Toledo}, {Aladro}, {Broguiere}, {Cortes}, {Cortes}, {Espada},
  {Galarza}, {Garcia-Appadoo}, {Guzman-Ramirez}, {Humphreys}, {Jung}, {Kameno},
  {Laing}, {Leon}, {Marconi}, {Mignano}, {Nikolic}, {Nyman}, {Radiszcz},
  {Remijan}, {Rod{\'o}n}, {Sawada}, {Takahashi}, {Tilanus}, {Vila Vilaro},
  {Watson}, {Wiklind}, {Akiyama}, {Chapillon}, {de Gregorio-Monsalvo}, {Di
  Francesco}, {Gueth}, {Kawamura}, {Lee}, {Nguyen Luong}, {Mangum}, {Pietu},
  {Sanhueza}, {Saigo}, {Takakuwa}, {Ubach}, {van Kempen}, {Wootten},
  {Castro-Carrizo}, {Francke}, {Gallardo}, {Garcia}, {Gonzalez}, {Hill},
  {Kaminski}, {Kurono}, {Liu}, {Lopez}, {Morales}, {Plarre}, {Schieven},
  {Testi}, {Videla}, {Villard}, {Andreani}, {Hibbard}, \&
  {Tatematsu}}]{ALMAPartnership2015}
{ALMA Partnership}, {Brogan}, C.~L., {P{\'e}rez}, L.~M., {et~al.} 2015, \apjl,
  808, L3

\bibitem[{{Amara} \& {Quanz}(2012)}]{Amara2012}
{Amara}, A. \& {Quanz}, S.~P. 2012, \mnras, 427, 948

\bibitem[{{Andrews} {et~al.}(2018){Andrews}, {Huang}, {P{\'e}rez}, {Isella},
  {Dullemond}, {Kurtovic}, {Guzm{\'a}n}, {Carpenter}, {Wilner}, {Zhang}, {Zhu},
  {Birnstiel}, {Bai}, {Benisty}, {Hughes}, {{\"O}berg}, \&
  {Ricci}}]{DSHARP_I_Andrews2018}
{Andrews}, S.~M., {Huang}, J., {P{\'e}rez}, L.~M., {et~al.} 2018, \apjl, 869,
  L41

\bibitem[{{Andrews} {et~al.}(2016){Andrews}, {Wilner}, {Zhu}, {Birnstiel},
  {Carpenter}, {P{\'e}rez}, {Bai}, {{\"O}berg}, {Hughes}, {Isella}, \&
  {Ricci}}]{Andrews2016}
{Andrews}, S.~M., {Wilner}, D.~J., {Zhu}, Z., {et~al.} 2016, \apjl, 820, L40

\bibitem[{{Antichi} {et~al.}(2009){Antichi}, {Dohlen}, {Gratton}, {Mesa},
  {Claudi}, {Giro}, {Boccaletti}, {Mouillet}, {Puget}, \&
  {Beuzit}}]{antichi2009bigre}
{Antichi}, J., {Dohlen}, K., {Gratton}, R.~G., {et~al.} 2009, \apj, 695, 1042

\bibitem[{{Aoyama} \& {Ikoma}(2019)}]{Aoyama2019}
{Aoyama}, Y. \& {Ikoma}, M. 2019, \apjl, 885, L29

\bibitem[{{Astropy Collaboration} {et~al.}(2013){Astropy Collaboration},
  {Robitaille}, {Tollerud}, {Greenfield}, {Droettboom}, {Bray}, {Aldcroft},
  {Davis}, {Ginsburg}, {Price-Whelan}, {Kerzendorf}, {Conley}, {Crighton},
  {Barbary}, {Muna}, {Ferguson}, {Grollier}, {Parikh}, {Nair}, {Unther},
  {Deil}, {Woillez}, {Conseil}, {Kramer}, {Turner}, {Singer}, {Fox}, {Weaver},
  {Zabalza}, {Edwards}, {Azalee Bostroem}, {Burke}, {Casey}, {Crawford},
  {Dencheva}, {Ely}, {Jenness}, {Labrie}, {Lim}, {Pierfederici}, {Pontzen},
  {Ptak}, {Refsdal}, {Servillat}, \& {Streicher}}]{astropy:2013}
{Astropy Collaboration}, {Robitaille}, T.~P., {Tollerud}, E.~J., {et~al.} 2013,
  \aap, 558, A33

\bibitem[{{Bacon} {et~al.}(2010){Bacon}, {Accardo}, {Adjali}, {Anwand},
  {Bauer}, {Biswas}, {Blaizot}, {Boudon}, {Brau-Nogue}, {Brinchmann},
  {Caillier}, {Capoani}, {Carollo}, {Contini}, {Couderc}, {Daguis{\'e}},
  {Deiries}, {Delabre}, {Dreizler}, {Dubois}, {Dupieux}, {Dupuy}, {Emsellem},
  {Fechner}, {Fleischmann}, {Fran{\c c}ois}, {Gallou}, {Gharsa}, {Glindemann},
  {Gojak}, {Guiderdoni}, {Hansali}, {Hahn}, {Jarno}, {Kelz}, {Koehler},
  {Kosmalski}, {Laurent}, {Le Floch}, {Lilly}, {Lizon}, {Loupias}, {Manescau},
  {Monstein}, {Nicklas}, {Olaya}, {Pares}, {Pasquini}, {P{\'e}contal-Rousset},
  {Pell{\'o}}, {Petit}, {Popow}, {Reiss}, {Remillieux}, {Renault}, {Roth},
  {Rupprecht}, {Serre}, {Schaye}, {Soucail}, {Steinmetz}, {Streicher}, {Stuik},
  {Valentin}, {Vernet}, {Weilbacher}, {Wisotzki}, \& {Yerle}}]{Bacon2010SPIE}
{Bacon}, R., {Accardo}, M., {Adjali}, L., {et~al.} 2010, in procspie, Vol.
  7735, Ground-based and Airborne Instrumentation for Astronomy III, 773508

\bibitem[{{Bacon} {et~al.}(2015){Bacon}, {Brinchmann}, {Richard}, {Contini},
  {Drake}, {Franx}, {Tacchella}, {Vernet}, {Wisotzki}, {Blaizot}, {Bouch{\'e}},
  {Bouwens}, {Cantalupo}, {Carollo}, {Carton}, {Caruana}, {Cl{\'e}ment},
  {Dreizler}, {Epinat}, {Guiderdoni}, {Herenz}, {Husser}, {Kamann}, {Kerutt},
  {Kollatschny}, {Krajnovic}, {Lilly}, {Martinsson}, {Michel-Dansac},
  {Patricio}, {Schaye}, {Shirazi}, {Soto}, {Soucail}, {Steinmetz}, {Urrutia},
  {Weilbacher}, \& {de Zeeuw}}]{2015A&A...575A..75B}
{Bacon}, R., {Brinchmann}, J., {Richard}, J., {et~al.} 2015, \aap, 575, A75

\bibitem[{{Bacon} {et~al.}(2017){Bacon}, {Conseil}, {Mary}, {Brinchmann},
  {Shepherd}, {Akhlaghi}, {Weilbacher}, {Piqueras}, {Wisotzki}, {Lagattuta},
  {Epinat}, {Guerou}, {Inami}, {Cantalupo}, {Courbot}, {Contini}, {Richard},
  {Maseda}, {Bouwens}, {Bouch{\'e}}, {Kollatschny}, {Schaye}, {Marino},
  {Pello}, {Herenz}, {Guiderdoni}, \& {Carollo}}]{2017A&A...608A...1B}
{Bacon}, R., {Conseil}, S., {Mary}, D., {et~al.} 2017, \aap, 608, A1

\bibitem[{{Benisty} {et~al.}(2015){Benisty}, {Juhasz}, {Boccaletti},
  {Avenhaus}, {Milli}, {Thalmann}, {Dominik}, {Pinilla}, {Buenzli}, {Pohl},
  {Beuzit}, {Birnstiel}, {de Boer}, {Bonnefoy}, {Chauvin}, {Christiaens},
  {Garufi}, {Grady}, {Henning}, {Huelamo}, {Isella}, {Langlois}, {M{\'e}nard},
  {Mouillet}, {Olofsson}, {Pantin}, {Pinte}, \& {Pueyo}}]{Benisty2015}
{Benisty}, M., {Juhasz}, A., {Boccaletti}, A., {et~al.} 2015, \aap, 578, L6

\bibitem[{{Beuzit} {et~al.}(2019){Beuzit}, {Vigan}, {Mouillet}, {Dohlen},
  {Gratton}, {Boccaletti}, {Sauvage}, {Schmid}, {Langlois}, {Petit},
  {Baruffolo}, {Feldt}, {Milli}, {Wahhaj}, {Abe}, {Anselmi}, {Antichi},
  {Barette}, {Baudrand}, {Baudoz}, {Bazzon}, {Bernardi}, {Blanchard}, {Brast},
  {Bruno}, {Buey}, {Carbillet}, {Carle}, {Cascone}, {Chapron}, {Charton},
  {Chauvin}, {Claudi}, {Costille}, {De Caprio}, {de Boer}, {Delboulb{\'e}},
  {Desidera}, {Dominik}, {Downing}, {Dupuis}, {Fabron}, {Fantinel}, {Farisato},
  {Feautrier}, {Fedrigo}, {Fusco}, {Gigan}, {Ginski}, {Girard}, {Giro},
  {Gisler}, {Gluck}, {Gry}, {Henning}, {Hubin}, {Hugot}, {Incorvaia}, {Jaquet},
  {Kasper}, {Lagadec}, {Lagrange}, {Le Coroller}, {Le Mignant}, {Le Ruyet},
  {Lessio}, {Lizon}, {Llored}, {Lundin}, {Madec}, {Magnard}, {Marteaud},
  {Martinez}, {Maurel}, {M{\'e}nard}, {Mesa}, {M{\"o}ller-Nilsson}, {Moulin},
  {Moutou}, {Orign{\'e}}, {Parisot}, {Pavlov}, {Perret}, {Pragt}, {Puget},
  {Rabou}, {Ramos}, {Reess}, {Rigal}, {Rochat}, {Roelfsema}, {Rousset}, {Roux},
  {Saisse}, {Salasnich}, {Santambrogio}, {Scuderi}, {Segransan}, {Sevin},
  {Siebenmorgen}, {Soenke}, {Stadler}, {Suarez}, {Tiph{\`e}ne}, {Turatto},
  {Udry}, {Vakili}, {Waters}, {Weber}, {Wildi}, {Zins}, \&
  {Zurlo}}]{Beuzit2019}
{Beuzit}, J.~L., {Vigan}, A., {Mouillet}, D., {et~al.} 2019, \aap, 631, A155

\bibitem[{{Biller} {et~al.}(2013){Biller}, {Liu}, {Wahhaj}, {Nielsen},
  {Hayward}, {Males}, {Skemer}, {Close}, {Chun}, {Ftaclas}, {Clarke}, {Thatte},
  {Shkolnik}, {Reid}, {Hartung}, {Boss}, {Lin}, {Alencar}, {de Gouveia Dal
  Pino}, {Gregorio-Hetem}, \& {Toomey}}]{Biller2013}
{Biller}, B.~A., {Liu}, M.~C., {Wahhaj}, Z., {et~al.} 2013, \apj, 777, 160

\bibitem[{{Biller} {et~al.}(2014){Biller}, {Males}, {Rodigas}, {Morzinski},
  {Close}, {Juh{\'a}sz}, {Follette}, {Lacour}, {Benisty}, {Sicilia-Aguilar},
  {Hinz}, {Weinberger}, {Henning}, {Pott}, {Bonnefoy}, \&
  {K{\"o}hler}}]{Biller2014}
{Biller}, B.~A., {Males}, J., {Rodigas}, T., {et~al.} 2014, \apjl, 792, L22

\bibitem[{{Brittain} {et~al.}(2014){Brittain}, {Carr}, {Najita}, {Quanz}, \&
  {Meyer}}]{Brittain2014}
{Brittain}, S.~D., {Carr}, J.~S., {Najita}, J.~R., {Quanz}, S.~P., \& {Meyer},
  M.~R. 2014, \apj, 791, 136

\bibitem[{{Brittain} {et~al.}(2013){Brittain}, {Najita}, {Carr}, {Liskowsky},
  {Troutman}, \& {Doppmann}}]{Brittain2013}
{Brittain}, S.~D., {Najita}, J.~R., {Carr}, J.~S., {et~al.} 2013, \apj, 767,
  159

\bibitem[{{Calvet} \& {Gullbring}(1998)}]{Calvet1998}
{Calvet}, N. \& {Gullbring}, E. 1998, \apj, 509, 802

\bibitem[{{Close} {et~al.}(2014){Close}, {Follette}, {Males}, {Puglisi},
  {Xompero}, {Apai}, {Najita}, {Weinberger}, {Morzinski}, {Rodigas}, {Hinz},
  {Bailey}, \& {Briguglio}}]{Close2014}
{Close}, L.~M., {Follette}, K.~B., {Males}, J.~R., {et~al.} 2014, \apjl, 781,
  L30

\bibitem[{{Cresci} {et~al.}(2015){Cresci}, {Marconi}, {Zibetti}, {Risaliti},
  {Carniani}, {Mannucci}, {Gallazzi}, {Maiolino}, {Balmaverde}, {Brusa},
  {Capetti}, {Cicone}, {Feruglio}, {Bland-Hawthorn}, {Nagao}, {Oliva},
  {Salvato}, {Sani}, {Tozzi}, {Urrutia}, \& {Venturi}}]{2015A&A...582A..63C}
{Cresci}, G., {Marconi}, A., {Zibetti}, S., {et~al.} 2015, \aap, 582, A63

\bibitem[{{Cugno} {et~al.}(2019){Cugno}, {Quanz}, {Hunziker}, {Stolker},
  {Schmid}, {Avenhaus}, {Baudoz}, {Bohn}, {Bonnefoy}, {Buenzli}, {Chauvin},
  {Cheetham}, {Desidera}, {Dominik}, {Feautrier}, {Feldt}, {Ginski}, {Girard},
  {Gratton}, {Hagelberg}, {Hugot}, {Janson}, {Lagrange}, {Langlois}, {Magnard},
  {Maire}, {Menard}, {Meyer}, {Milli}, {Mordasini}, {Pinte}, {Pragt},
  {Roelfsema}, {Rigal}, {Szul{\'a}gyi}, {van Boekel}, {van der Plas}, {Vigan},
  {Wahhaj}, \& {Zurlo}}]{Cugno2019}
{Cugno}, G., {Quanz}, S.~P., {Hunziker}, S., {et~al.} 2019, \aap, 622, A156

\bibitem[{{Currie} {et~al.}(2015){Currie}, {Cloutier}, {Brittain}, {Grady},
  {Burrows}, {Muto}, {Kenyon}, \& {Kuchner}}]{Currie2015}
{Currie}, T., {Cloutier}, R., {Brittain}, S., {et~al.} 2015, \apjl, 814, L27

\bibitem[{{Currie} {et~al.}(2019){Currie}, {Marois}, {Cieza}, {Mulders},
  {Lawson}, {Caceres}, {Rodriguez-Ruiz}, {Wisniewski}, {Guyon}, {Brandt},
  {Kasdin}, {Groff}, {Lozi}, {Chilcote}, {Hodapp}, {Jovanovic}, {Martinache},
  {Skaf}, {Lyra}, {Tamura}, {Asensio-Torres}, {Dong}, {Grady}, {Gerard},
  {Fukagawa}, {Hand}, {Hayashi}, {Henning}, {Kudo}, {Kuzuhara}, {Kwon},
  {McElwain}, \& {Uyama}}]{Currie2019}
{Currie}, T., {Marois}, C., {Cieza}, L., {et~al.} 2019, \apjl, 877, L3

\bibitem[{{de Juan Ovelar} {et~al.}(2013){de Juan Ovelar}, {Min}, {Dominik},
  {Thalmann}, {Pinilla}, {Benisty}, \& {Birnstiel}}]{deJuanOvelar2013}
{de Juan Ovelar}, M., {Min}, M., {Dominik}, C., {et~al.} 2013, \aap, 560, A111

\bibitem[{{Dong} \& {Fung}(2017)}]{DongFung2017}
{Dong}, R. \& {Fung}, J. 2017, \apj, 835, 146

\bibitem[{{Dong} {et~al.}(2015{\natexlab{a}}){Dong}, {Zhu}, {Rafikov}, \&
  {Stone}}]{Dong2015a}
{Dong}, R., {Zhu}, Z., {Rafikov}, R.~R., \& {Stone}, J.~M. 2015{\natexlab{a}},
  \apjl, 809, L5

\bibitem[{{Dong} {et~al.}(2015{\natexlab{b}}){Dong}, {Zhu}, \&
  {Whitney}}]{Dong2015b}
{Dong}, R., {Zhu}, Z., \& {Whitney}, B. 2015{\natexlab{b}}, \apj, 809, 93

\bibitem[{{Ellerbroek} {et~al.}(2014){Ellerbroek}, {Podio}, {Dougados},
  {Cabrit}, {Sitko}, {Sana}, {Kaper}, {de Koter}, {Klaassen}, {Mulders},
  {Mendigut{\'\i}a}, {Grady}, {Grankin}, {van Winckel}, {Bacciotti}, {Russell},
  {Lynch}, {Hammel}, {Beerman}, {Day}, {Huelsman}, {Werren}, {Henden}, \&
  {Grindlay}}]{Ellerbroek2014}
{Ellerbroek}, L.~E., {Podio}, L., {Dougados}, C., {et~al.} 2014, \aap, 563, A87

\bibitem[{{Fang} {et~al.}(2009){Fang}, {van Boekel}, {Wang}, {Carmona},
  {Sicilia-Aguilar}, \& {Henning}}]{Fang2009}
{Fang}, M., {van Boekel}, R., {Wang}, W., {et~al.} 2009, \aap, 504, 461

\bibitem[{{Flock} {et~al.}(2015){Flock}, {Ruge}, {Dzyurkevich}, {Henning},
  {Klahr}, \& {Wolf}}]{Flock2015}
{Flock}, M., {Ruge}, J.~P., {Dzyurkevich}, N., {et~al.} 2015, \aap, 574, A68

\bibitem[{{Follette} {et~al.}(2017){Follette}, {Rameau}, {Dong}, {Pueyo},
  {Close}, {Duch{\^e}ne}, {Fung}, {Leonard}, {Macintosh}, {Males}, {Marois},
  {Millar-Blanchaer}, {Morzinski}, {Mullen}, {Perrin}, {Spiro}, {Wang},
  {Ammons}, {Bailey}, {Barman}, {Bulger}, {Chilcote}, {Cotten}, {De Rosa},
  {Doyon}, {Fitzgerald}, {Goodsell}, {Graham}, {Greenbaum}, {Hibon}, {Hung},
  {Ingraham}, {Kalas}, {Konopacky}, {Larkin}, {Maire}, {Marchis}, {Metchev},
  {Nielsen}, {Oppenheimer}, {Palmer}, {Patience}, {Poyneer}, {Rajan},
  {Rantakyr{\"o}}, {Savransky}, {Schneider}, {Sivaramakrishnan}, {Song},
  {Soummer}, {Thomas}, {Vega}, {Wallace}, {Ward-Duong}, {Wiktorowicz}, \&
  {Wolff}}]{Follette2017}
{Follette}, K.~B., {Rameau}, J., {Dong}, R., {et~al.} 2017, \aj, 153, 264

\bibitem[{{Fung} \& {Dong}(2015)}]{FungDong2015}
{Fung}, J. \& {Dong}, R. 2015, \apjl, 815, L21

\bibitem[{{Gaia Collaboration} {et~al.}(2018){Gaia Collaboration}, {Brown},
  {Vallenari}, {Prusti}, {de Bruijne}, {Babusiaux}, {Bailer-Jones}, {Biermann},
  {Evans}, {Eyer}, {Jansen}, {Jordi}, {Klioner}, {Lammers}, {Lindegren},
  {Luri}, {Mignard}, {Panem}, {Pourbaix}, {Randich}, {Sartoretti}, {Siddiqui},
  {Soubiran}, {van Leeuwen}, {Walton}, {Arenou}, {Bastian}, {Cropper},
  {Drimmel}, {Katz}, {Lattanzi}, {Bakker}, {Cacciari}, {Casta{\~n}eda},
  {Chaoul}, {Cheek}, {De Angeli}, {Fabricius}, {Guerra}, {Holl}, {Masana},
  {Messineo}, {Mowlavi}, {Nienartowicz}, {Panuzzo}, {Portell}, {Riello},
  {Seabroke}, {Tanga}, {Th{\'e}venin}, {Gracia-Abril}, {Comoretto},
  {Garcia-Reinaldos}, {Teyssier}, {Altmann}, {Andrae}, {Audard},
  {Bellas-Velidis}, {Benson}, {Berthier}, {Blomme}, {Burgess}, {Busso},
  {Carry}, {Cellino}, {Clementini}, {Clotet}, {Creevey}, {Davidson}, {De
  Ridder}, {Delchambre}, {Dell'Oro}, {Ducourant},
  {Fern{\'a}ndez-Hern{\'a}ndez}, {Fouesneau}, {Fr{\'e}mat}, {Galluccio},
  {Garc{\'\i}a-Torres}, {Gonz{\'a}lez-N{\'u}{\~n}ez}, {Gonz{\'a}lez-Vidal},
  {Gosset}, {Guy}, {Halbwachs}, {Hambly}, {Harrison}, {Hern{\'a}ndez},
  {Hestroffer}, {Hodgkin}, {Hutton}, {Jasniewicz}, {Jean-Antoine-Piccolo},
  {Jordan}, {Korn}, {Krone-Martins}, {Lanzafame}, {Lebzelter}, {L{\"o}ffler},
  {Manteiga}, {Marrese}, {Mart{\'\i}n-Fleitas}, {Moitinho}, {Mora}, {Muinonen},
  {Osinde}, {Pancino}, {Pauwels}, {Petit}, {Recio-Blanco}, {Richards},
  {Rimoldini}, {Robin}, {Sarro}, {Siopis}, {Smith}, {Sozzetti}, {S{\"u}veges},
  {Torra}, {van Reeven}, {Abbas}, {Abreu Aramburu}, {Accart}, {Aerts},
  {Altavilla}, {{\'A}lvarez}, {Alvarez}, {Alves}, {Anderson}, {Andrei},
  {Anglada Varela}, {Antiche}, {Antoja}, {Arcay}, {Astraatmadja}, {Bach},
  {Baker}, {Balaguer-N{\'u}{\~n}ez}, {Balm}, {Barache}, {Barata}, {Barbato},
  {Barblan}, {Barklem}, {Barrado}, {Barros}, {Barstow}, {Bartholom{\'e}
  Mu{\~n}oz}, {Bassilana}, {Becciani}, {Bellazzini}, {Berihuete}, {Bertone},
  {Bianchi}, {Bienaym{\'e}}, {Blanco-Cuaresma}, {Boch}, {Boeche}, {Bombrun},
  {Borrachero}, {Bossini}, {Bouquillon}, {Bourda}, {Bragaglia}, {Bramante},
  {Breddels}, {Bressan}, {Brouillet}, {Br{\"u}semeister}, {Brugaletta},
  {Bucciarelli}, {Burlacu}, {Busonero}, {Butkevich}, {Buzzi}, {Caffau},
  {Cancelliere}, {Cannizzaro}, {Cantat-Gaudin}, {Carballo}, {Carlucci},
  {Carrasco}, {Casamiquela}, {Castellani}, {Castro-Ginard}, {Charlot},
  {Chemin}, {Chiavassa}, {Cocozza}, {Costigan}, {Cowell}, {Crifo}, {Crosta},
  {Crowley}, {Cuypers}, {Dafonte}, {Damerdji}, {Dapergolas}, {David}, {David},
  {de Laverny}, {De Luise}, {De March}, {de Martino}, {de Souza}, {de Torres},
  {Debosscher}, {del Pozo}, {Delbo}, {Delgado}, {Delgado}, {Di Matteo},
  {Diakite}, {Diener}, {Distefano}, {Dolding}, {Drazinos}, {Dur{\'a}n},
  {Edvardsson}, {Enke}, {Eriksson}, {Esquej}, {Eynard Bontemps}, {Fabre},
  {Fabrizio}, {Faigler}, {Falc{\~a}o}, {Farr{\`a}s Casas}, {Federici},
  {Fedorets}, {Fernique}, {Figueras}, {Filippi}, {Findeisen}, {Fonti},
  {Fraile}, {Fraser}, {Fr{\'e}zouls}, {Gai}, {Galleti}, {Garabato},
  {Garc{\'\i}a-Sedano}, {Garofalo}, {Garralda}, {Gavel}, {Gavras}, {Gerssen},
  {Geyer}, {Giacobbe}, {Gilmore}, {Girona}, {Giuffrida}, {Glass}, {Gomes},
  {Granvik}, {Gueguen}, {Guerrier}, {Guiraud}, {Guti{\'e}rrez-S{\'a}nchez},
  {Haigron}, {Hatzidimitriou}, {Hauser}, {Haywood}, {Heiter}, {Helmi}, {Heu},
  {Hilger}, {Hobbs}, {Hofmann}, {Holland}, {Huckle}, {Hypki}, {Icardi},
  {Jan{\ss}en}, {Jevardat de Fombelle}, {Jonker}, {Juh{\'a}sz}, {Julbe},
  {Karampelas}, {Kewley}, {Klar}, {Kochoska}, {Kohley}, {Kolenberg},
  {Kontizas}, {Kontizas}, {Koposov}, {Kordopatis}, {Kostrzewa-Rutkowska},
  {Koubsky}, {Lambert}, {Lanza}, {Lasne}, {Lavigne}, {Le Fustec}, {Le
  Poncin-Lafitte}, {Lebreton}, {Leccia}, {Leclerc}, {Lecoeur-Taibi},
  {Lenhardt}, {Leroux}, {Liao}, {Licata}, {Lindstr{\o}m}, {Lister}, {Livanou},
  {Lobel}, {L{\'o}pez}, {Managau}, {Mann}, {Mantelet}, {Marchal}, {Marchant},
  {Marconi}, {Marinoni}, {Marschalk{\'o}}, {Marshall}, {Martino}, {Marton},
  {Mary}, {Massari}, {Matijevi{\v{c}}}, {Mazeh}, {McMillan}, {Messina},
  {Michalik}, {Millar}, {Molina}, {Molinaro}, {Moln{\'a}r}, {Montegriffo},
  {Mor}, {Morbidelli}, {Morel}, {Morris}, {Mulone}, {Muraveva}, {Musella},
  {Nelemans}, {Nicastro}, {Noval}, {O'Mullane}, {Ord{\'e}novic},
  {Ord{\'o}{\~n}ez-Blanco}, {Osborne}, {Pagani}, {Pagano}, {Pailler},
  {Palacin}, {Palaversa}, {Panahi}, {Pawlak}, {Piersimoni}, {Pineau}, {Plachy},
  {Plum}, {Poggio}, {Poujoulet}, {Pr{\v{s}}a}, {Pulone}, {Racero}, {Ragaini},
  {Rambaux}, {Ramos-Lerate}, {Regibo}, {Reyl{\'e}}, {Riclet}, {Ripepi}, {Riva},
  {Rivard}, {Rixon}, {Roegiers}, {Roelens}, {Romero-G{\'o}mez}, {Rowell},
  {Royer}, {Ruiz-Dern}, {Sadowski}, {Sagrist{\`a} Sell{\'e}s}, {Sahlmann},
  {Salgado}, {Salguero}, {Sanna}, {Santana-Ros}, {Sarasso}, {Savietto},
  {Schultheis}, {Sciacca}, {Segol}, {Segovia}, {S{\'e}gransan}, {Shih},
  {Siltala}, {Silva}, {Smart}, {Smith}, {Solano}, {Solitro}, {Sordo}, {Soria
  Nieto}, {Souchay}, {Spagna}, {Spoto}, {Stampa}, {Steele},
  {Steidelm{\"u}ller}, {Stephenson}, {Stoev}, {Suess}, {Surdej}, {Szabados},
  {Szegedi-Elek}, {Tapiador}, {Taris}, {Tauran}, {Taylor}, {Teixeira},
  {Terrett}, {Teyssand ier}, {Thuillot}, {Titarenko}, {Torra Clotet}, {Turon},
  {Ulla}, {Utrilla}, {Uzzi}, {Vaillant}, {Valentini}, {Valette}, {van Elteren},
  {Van Hemelryck}, {van Leeuwen}, {Vaschetto}, {Vecchiato}, {Veljanoski},
  {Viala}, {Vicente}, {Vogt}, {von Essen}, {Voss}, {Votruba}, {Voutsinas},
  {Walmsley}, {Weiler}, {Wertz}, {Wevers}, {Wyrzykowski}, {Yoldas},
  {{\v{Z}}erjal}, {Ziaeepour}, {Zorec}, {Zschocke}, {Zucker}, {Zurbach}, \&
  {Zwitter}}]{GaiaCollaboration2018}
{Gaia Collaboration}, {Brown}, A.~G.~A., {Vallenari}, A., {et~al.} 2018, \aap,
  616, A1

\bibitem[{{Garufi} {et~al.}(2013){Garufi}, {Quanz}, {Avenhaus}, {Buenzli},
  {Dominik}, {Meru}, {Meyer}, {Pinilla}, {Schmid}, \& {Wolf}}]{Garufi2013}
{Garufi}, A., {Quanz}, S.~P., {Avenhaus}, H., {et~al.} 2013, \aap, 560, A105

\bibitem[{{Guimar{\~a}es} {et~al.}(2006){Guimar{\~a}es}, {Alencar}, {Corradi},
  \& {Vieira}}]{Guimaraes2006}
{Guimar{\~a}es}, M.~M., {Alencar}, S.~H.~P., {Corradi}, W.~J.~B., \& {Vieira},
  S.~L.~A. 2006, \aap, 457, 581

\bibitem[{{Gullbring} {et~al.}(1998){Gullbring}, {Hartmann}, {Brice{\~n}o}, \&
  {Calvet}}]{Gullbring1998}
{Gullbring}, E., {Hartmann}, L., {Brice{\~n}o}, C., \& {Calvet}, N. 1998, \apj,
  492, 323

\bibitem[{{Haffert} {et~al.}(2019){Haffert}, {Bohn}, {de Boer}, {Snellen},
  {Brinchmann}, {Girard}, {Keller}, \& {Bacon}}]{Haffert2019}
{Haffert}, S.~Y., {Bohn}, A.~J., {de Boer}, J., {et~al.} 2019, Nature
  Astronomy, 3, 749

\bibitem[{{Hashimoto} {et~al.}(2020){Hashimoto}, {Aoyama}, {Konishi}, {Uyama},
  {Takasao}, {Ikoma}, \& {Tanigawa}}]{Hashimoto2020}
{Hashimoto}, J., {Aoyama}, Y., {Konishi}, M., {et~al.} 2020, \aj, 159, 222

\bibitem[{{Herczeg} \& {Hillenbrand}(2008)}]{Herczeg2008}
{Herczeg}, G.~J. \& {Hillenbrand}, L.~A. 2008, \apj, 681, 594

\bibitem[{{Hoeijmakers} {et~al.}(2018){Hoeijmakers}, {Schwarz}, {Snellen}, {de
  Kok}, {Bonnefoy}, {Chauvin}, {Lagrange}, \& {Girard}}]{Hoeijmakers2018}
{Hoeijmakers}, H.~J., {Schwarz}, H., {Snellen}, I.~A.~G., {et~al.} 2018, \aap,
  617, A144

\bibitem[{{Hu{\'e}lamo} {et~al.}(2018){Hu{\'e}lamo}, {Chauvin}, {Schmid},
  {Quanz}, {Whelan}, {Lillo-Box}, {Barrado}, {Montesinos}, {Alcal{\'a}},
  {Benisty}, {de Gregorio-Monsalvo}, {Mendigut{\'\i}a}, {Bouy}, {Mer{\'\i}n},
  {de Boer}, {Garufi}, \& {Pantin}}]{Huelamo2018}
{Hu{\'e}lamo}, N., {Chauvin}, G., {Schmid}, H.~M., {et~al.} 2018, \aap, 613, L5

\bibitem[{{Husser} {et~al.}(2016){Husser}, {Kamann}, {Dreizler}, {Wendt},
  {Wulff}, {Bacon}, {Wisotzki}, {Brinchmann}, {Weilbacher}, {Roth}, \&
  {Monreal-Ibero}}]{2016A&A...588A.148H}
{Husser}, T.-O., {Kamann}, S., {Dreizler}, S., {et~al.} 2016, \aap, 588, A148

\bibitem[{{Isella} {et~al.}(2019){Isella}, {Benisty}, {Teague}, {Bae},
  {Keppler}, {Facchini}, \& {P{\'e}rez}}]{Isella2019}
{Isella}, A., {Benisty}, M., {Teague}, R., {et~al.} 2019, \apjl, 879, L25

\bibitem[{{Isella} {et~al.}(2018){Isella}, {Huang}, {Andrews}, {Dullemond},
  {Birnstiel}, {Zhang}, {Zhu}, {Guzm{\'a}n}, {P{\'e}rez}, {Bai}, {Benisty},
  {Carpenter}, {Ricci}, \& {Wilner}}]{DSHARP_IX_Isella2018}
{Isella}, A., {Huang}, J., {Andrews}, S.~M., {et~al.} 2018, \apjl, 869, L49

\bibitem[{{Kamann} {et~al.}(2018){Kamann}, {Husser}, {Dreizler}, {Emsellem},
  {Weilbacher}, {Martens}, {Bacon}, {den Brok}, {Giesers}, {Krajnovi{\'c}},
  {Roth}, {Wendt}, \& {Wisotzki}}]{2018MNRAS.473.5591K}
{Kamann}, S., {Husser}, T.~O., {Dreizler}, S., {et~al.} 2018, \mnras, 473, 5591

\bibitem[{{Keppler} {et~al.}(2018){Keppler}, {Benisty}, {M{\"u}ller},
  {Henning}, {van Boekel}, {Cantalloube}, {Ginski}, {van Holstein}, {Maire},
  {Pohl}, {Samland }, {Avenhaus}, {Baudino}, {Boccaletti}, {de Boer},
  {Bonnefoy}, {Chauvin}, {Desidera}, {Langlois}, {Lazzoni}, {Marleau},
  {Mordasini}, {Pawellek}, {Stolker}, {Vigan}, {Zurlo}, {Birnstiel},
  {Brandner}, {Feldt}, {Flock}, {Girard}, {Gratton}, {Hagelberg}, {Isella},
  {Janson}, {Juhasz}, {Kemmer}, {Kral}, {Lagrange}, {Launhardt}, {Matter},
  {M{\'e}nard}, {Milli}, {Molli{\`e}re}, {Olofsson}, {P{\'e}rez}, {Pinilla},
  {Pinte}, {Quanz}, {Schmidt}, {Udry}, {Wahhaj}, {Williams}, {Buenzli},
  {Cudel}, {Dominik}, {Galicher}, {Kasper}, {Lannier}, {Mesa}, {Mouillet},
  {Peretti}, {Perrot}, {Salter}, {Sissa}, {Wildi}, {Abe}, {Antichi},
  {Augereau}, {Baruffolo}, {Baudoz}, {Bazzon}, {Beuzit}, {Blanchard}, {Brems},
  {Buey}, {De Caprio}, {Carbillet}, {Carle}, {Cascone}, {Cheetham}, {Claudi},
  {Costille}, {Delboulb{\'e}}, {Dohlen}, {Fantinel}, {Feautrier}, {Fusco},
  {Giro}, {Gluck}, {Gry}, {Hubin}, {Hugot}, {Jaquet}, {Le Mignant}, {Llored},
  {Madec}, {Magnard}, {Martinez}, {Maurel}, {Meyer}, {M{\"o}ller-Nilsson},
  {Moulin}, {Mugnier}, {Orign{\'e}}, {Pavlov}, {Perret}, {Petit}, {Pragt},
  {Puget}, {Rabou}, {Ramos}, {Rigal}, {Rochat}, {Roelfsema}, {Rousset}, {Roux},
  {Salasnich}, {Sauvage}, {Sevin}, {Soenke}, {Stadler}, {Suarez}, {Turatto}, \&
  {Weber}}]{Keppler2018}
{Keppler}, M., {Benisty}, M., {M{\"u}ller}, A., {et~al.} 2018, \aap, 617, A44

\bibitem[{{Kley} \& {Nelson}(2012)}]{KleyNelson2012}
{Kley}, W. \& {Nelson}, R.~P. 2012, \araa, 50, 211

\bibitem[{{Kraus} \& {Ireland}(2012)}]{Kraus2012}
{Kraus}, A.~L. \& {Ireland}, M.~J. 2012, \apj, 745, 5

\bibitem[{{Kuhn} {et~al.}(2001){Kuhn}, {Potter}, \& {Parise}}]{Kuhn2001_PDI}
{Kuhn}, J.~R., {Potter}, D., \& {Parise}, B. 2001, \apjl, 553, L189

\bibitem[{{L{\'e}pine} \& {Simon}(2009)}]{2009AJ....137.3632L}
{L{\'e}pine}, S. \& {Simon}, M. 2009, \aj, 137, 3632

\bibitem[{{Ligi} {et~al.}(2018){Ligi}, {Vigan}, {Gratton}, {de Boer},
  {Benisty}, {Boccaletti}, {Quanz}, {Meyer}, {Ginski}, {Sissa}, {Gry},
  {Henning}, {Beuzit}, {Biller}, {Bonnefoy}, {Chauvin}, {Cheetham}, {Cudel},
  {Delorme}, {Desidera}, {Feldt}, {Galicher}, {Girard}, {Janson}, {Kasper},
  {Kopytova}, {Lagrange}, {Langlois}, {Lecoroller}, {Maire}, {M{\'e}nard},
  {Mesa}, {Peretti}, {Perrot}, {Pinilla}, {Pohl}, {Rouan}, {Stolker},
  {Samland}, {Wahhaj}, {Wildi}, {Zurlo}, {Buey}, {Fantinel}, {Fusco}, {Jaquet},
  {Moulin}, {Ramos}, {Suarez}, \& {Weber}}]{Ligi2018}
{Ligi}, R., {Vigan}, A., {Gratton}, R., {et~al.} 2018, \mnras, 473, 1774

\bibitem[{{Madec} {et~al.}(2018){Madec}, {Arsenault}, {Kuntschner}, {Kolb},
  {Pirard}, {Paufique}, {La Penna}, {Hackenberg}, {Vernet}, {Su{\'a}rez
  Valles}, \& {Hubin}}]{2018SPIE10703E..02M}
{Madec}, P.-Y., {Arsenault}, R., {Kuntschner}, H., {et~al.} 2018, in Society of
  Photo-Optical Instrumentation Engineers (SPIE) Conference Series, Vol. 10703,
  Adaptive Optics Systems VI, 1070302

\bibitem[{{Marleau} {et~al.}(2017){Marleau}, {Klahr}, {Kuiper}, \&
  {Mordasini}}]{Marleau2017}
{Marleau}, G.-D., {Klahr}, H., {Kuiper}, R., \& {Mordasini}, C. 2017, \apj,
  836, 221

\bibitem[{{Marois} {et~al.}(2006){Marois}, {Lafreni{\`e}re}, {Doyon},
  {Macintosh}, \& {Nadeau}}]{Marois2006_ADI}
{Marois}, C., {Lafreni{\`e}re}, D., {Doyon}, R., {Macintosh}, B., \& {Nadeau},
  D. 2006, \apj, 641, 556

\bibitem[{{Mawet} {et~al.}(2014){Mawet}, {Milli}, {Wahhaj}, {Pelat}, {Absil},
  {Delacroix}, {Boccaletti}, {Kasper}, {Kenworthy}, {Marois}, {Mennesson}, \&
  {Pueyo}}]{Mawet2014}
{Mawet}, D., {Milli}, J., {Wahhaj}, Z., {et~al.} 2014, \apj, 792, 97

\bibitem[{{Mesa} {et~al.}(2019{\natexlab{a}}){Mesa}, {Keppler}, {Cantalloube},
  {Rodet}, {Charnay}, {Gratton}, {Langlois}, {Boccaletti}, {Bonnefoy}, {Vigan},
  {Flasseur}, {Bae}, {Benisty}, {Chauvin}, {de Boer}, {Desidera}, {Henning},
  {Lagrange}, {Meyer}, {Milli}, {M{\"u}ller}, {Pairet}, {Zurlo}, {Antoniucci},
  {Baudino}, {Brown Sevilla}, {Cascone}, {Cheetham}, {Claudi}, {Delorme},
  {D'Orazi}, {Feldt}, {Hagelberg}, {Janson}, {Kral}, {Lagadec}, {Lazzoni},
  {Ligi}, {Maire}, {Martinez}, {Menard}, {Meunier}, {Perrot}, {Petrus},
  {Pinte}, {Rickman}, {Rochat}, {Rouan}, {Samland}, {Sauvage}, {Schmidt},
  {Udry}, {Weber}, \& {Wildi}}]{Mesa2019b}
{Mesa}, D., {Keppler}, M., {Cantalloube}, F., {et~al.} 2019{\natexlab{a}},
  \aap, 632, A25

\bibitem[{{Mesa} {et~al.}(2019{\natexlab{b}}){Mesa}, {Langlois}, {Garufi},
  {Gratton}, {Desidera}, {D'Orazi}, {Flasseur}, {Barbieri}, {Benisty},
  {Henning}, {Ligi}, {Sissa}, {Vigan}, {Zurlo}, {Boccaletti}, {Bonnefoy},
  {Cantalloube}, {Chauvin}, {Cheetham}, {De Caprio}, {Delorme}, {Feldt},
  {Fusco}, {Gluck}, {Hagelberg}, {Lagrange}, {Lazzoni}, {Madec}, {Maire},
  {Menard}, {Meyer}, {Ramos}, {Rickman}, {Rouan}, {Schmidt}, \& {Van der
  Plas}}]{Mesa2019a}
{Mesa}, D., {Langlois}, M., {Garufi}, A., {et~al.} 2019{\natexlab{b}}, \mnras,
  488, 37

\bibitem[{{Messina} {et~al.}(2017){Messina}, {Lanzafame}, {Malo}, {Desidera},
  {Buccino}, {Zhang}, {Artemenko}, {Millward}, \& {Hambsch}}]{Messina2017}
{Messina}, S., {Lanzafame}, A.~C., {Malo}, L., {et~al.} 2017, \aap, 607, A3

\bibitem[{{Montesinos} {et~al.}(2009){Montesinos}, {Eiroa}, {Mora}, \&
  {Mer{\'\i}n}}]{Montesinos2009}
{Montesinos}, B., {Eiroa}, C., {Mora}, A., \& {Mer{\'\i}n}, B. 2009, \aap, 495,
  901

\bibitem[{{Mora} {et~al.}(2001){Mora}, {Mer{\'\i}n}, {Solano}, {Montesinos},
  {de Winter}, {Eiroa}, {Ferlet}, {Grady}, {Davies}, {Miranda}, {Oudmaijer},
  {Palacios}, {Quirrenbach}, {Harris}, {Rauer}, {Collier Cameron}, {Deeg},
  {Garz{\'o}n}, {Penny}, {Schneider}, {Tsapras}, \&
  {Wesselius}}]{2001A&A...378..116M}
{Mora}, A., {Mer{\'\i}n}, B., {Solano}, E., {et~al.} 2001, \aap, 378, 116

\bibitem[{{Mordasini} {et~al.}(2017){Mordasini}, {Marleau}, \&
  {Molli{\`e}re}}]{Mordasini2017}
{Mordasini}, C., {Marleau}, G.~D., \& {Molli{\`e}re}, P. 2017, \aap, 608, A72

\bibitem[{{M{\"u}ller} {et~al.}(2018){M{\"u}ller}, {Keppler}, {Henning},
  {Samland}, {Chauvin}, {Beust}, {Maire}, {Molaverdikhani}, {van Boekel},
  {Benisty}, {Boccaletti}, {Bonnefoy}, {Cantalloube}, {Charnay}, {Baudino},
  {Gennaro}, {Long}, {Cheetham}, {Desidera}, {Feldt}, {Fusco}, {Girard},
  {Gratton}, {Hagelberg}, {Janson}, {Lagrange}, {Langlois}, {Lazzoni}, {Ligi},
  {M{\'e}nard}, {Mesa}, {Meyer}, {Molli{\`e}re}, {Mordasini}, {Moulin},
  {Pavlov}, {Pawellek}, {Quanz}, {Ramos}, {Rouan}, {Sissa}, {Stadler}, {Vigan},
  {Wahhaj}, {Weber}, \& {Zurlo}}]{Muller2018}
{M{\"u}ller}, A., {Keppler}, M., {Henning}, T., {et~al.} 2018, \aap, 617, L2

\bibitem[{{Muro-Arena} {et~al.}(2019){Muro-Arena}, {Benisty}, {Ginski},
  {Dominik}, {Facchini}, {Villenave}, {van Boekel}, {Chauvin}, {Garufi},
  {Henning}, {Janson}, {Keppler}, {Matter}, {M{\'e}nard}, {Stolker}, {Zurlo},
  {Blanchard}, {Maurel}, {Moeller-Nilsson}, {Petit}, {Roux}, {Sevin}, \&
  {Wildi}}]{Muro-Arena2019}
{Muro-Arena}, G.~A., {Benisty}, M., {Ginski}, C., {et~al.} 2019, arXiv
  e-prints, arXiv:1911.09612

\bibitem[{{Muto} {et~al.}(2012){Muto}, {Grady}, {Hashimoto}, {Fukagawa},
  {Hornbeck}, {Sitko}, {Russell}, {Werren}, {Cur{\'e}}, {Currie}, {Ohashi},
  {Okamoto}, {Momose}, {Honda}, {Inutsuka}, {Takeuchi}, {Dong}, {Abe},
  {Brandner}, {Brandt}, {Carson}, {Egner}, {Feldt}, {Fukue}, {Goto}, {Guyon},
  {Hayano}, {Hayashi}, {Hayashi}, {Henning}, {Hodapp}, {Ishii}, {Iye},
  {Janson}, {Kandori}, {Knapp}, {Kudo}, {Kusakabe}, {Kuzuhara}, {Matsuo},
  {Mayama}, {McElwain}, {Miyama}, {Morino}, {Moro-Martin}, {Nishimura}, {Pyo},
  {Serabyn}, {Suto}, {Suzuki}, {Takami}, {Takato}, {Terada}, {Thalmann},
  {Tomono}, {Turner}, {Watanabe}, {Wisniewski}, {Yamada}, {Takami}, {Usuda}, \&
  {Tamura}}]{Muto2012}
{Muto}, T., {Grady}, C.~A., {Hashimoto}, J., {et~al.} 2012, \apjl, 748, L22

\bibitem[{{Natta} {et~al.}(2004){Natta}, {Testi}, {Muzerolle}, {Randich},
  {Comer{\'o}n}, \& {Persi}}]{Natta2004}
{Natta}, A., {Testi}, L., {Muzerolle}, J., {et~al.} 2004, \aap, 424, 603

\bibitem[{{Oberti} {et~al.}(2016){Oberti}, {Kolb}, {Le Louarn}, {La Penna},
  {Madec}, {Neichel}, {Sauvage}, {Fusco}, {Donaldson}, {Soenke}, {Su{\'a}rez
  Valles}, \& {Arsenault}}]{2016SPIE.9909E..1UO}
{Oberti}, S., {Kolb}, J., {Le Louarn}, M., {et~al.} 2016, in procspie, Vol.
  9909, Adaptive Optics Systems V, 99091U

\bibitem[{{Pecaut} \& {Mamajek}(2016)}]{PecautMamajek2016}
{Pecaut}, M.~J. \& {Mamajek}, E.~E. 2016, \mnras, 461, 794

\bibitem[{{Pineda} {et~al.}(2019){Pineda}, {Szul{\'a}gyi}, {Quanz}, {van
  Dishoeck}, {Garufi}, {Meru}, {Mulders}, {Testi}, {Meyer}, \&
  {Reggiani}}]{Pineda2019}
{Pineda}, J.~E., {Szul{\'a}gyi}, J., {Quanz}, S.~P., {et~al.} 2019, \apj, 871,
  48

\bibitem[{{Pinilla} {et~al.}(2012){Pinilla}, {Benisty}, \&
  {Birnstiel}}]{Pinilla2012}
{Pinilla}, P., {Benisty}, M., \& {Birnstiel}, T. 2012, \aap, 545, A81

\bibitem[{{Poggianti} {et~al.}(2017){Poggianti}, {Jaff{\'e}}, {Moretti},
  {Gullieuszik}, {Radovich}, {Tonnesen}, {Fritz}, {Bettoni}, {Vulcani},
  {Fasano}, {Bellhouse}, {Hau}, \& {Omizzolo}}]{2017Natur.548..304P}
{Poggianti}, B.~M., {Jaff{\'e}}, Y.~L., {Moretti}, A., {et~al.} 2017, \nat,
  548, 304

\bibitem[{{Price-Whelan} {et~al.}(2018){Price-Whelan}, {Sip{\H{o}}cz},
  {G{\"u}nther}, {Lim}, {Crawford}, {Conseil}, {Shupe}, {Craig}, {Dencheva},
  {Ginsburg}, {VanderPlas}, {Bradley}, {P{\'e}rez-Su{\'a}rez}, {de Val-Borro},
  {Paper Contributors}, {Aldcroft}, {Cruz}, {Robitaille}, {Tollerud},
  {Coordination Committee}, {Ardelean}, {Babej}, {Bach}, {Bachetti}, {Bakanov},
  {Bamford}, {Barentsen}, {Barmby}, {Baumbach}, {Berry}, {Biscani}, {Boquien},
  {Bostroem}, {Bouma}, {Brammer}, {Bray}, {Breytenbach}, {Buddelmeijer},
  {Burke}, {Calderone}, {Cano Rodr{\'\i}guez}, {Cara}, {Cardoso}, {Cheedella},
  {Copin}, {Corrales}, {Crichton}, {D{\textquoteright}Avella}, {Deil},
  {Depagne}, {Dietrich}, {Donath}, {Droettboom}, {Earl}, {Erben}, {Fabbro},
  {Ferreira}, {Finethy}, {Fox}, {Garrison}, {Gibbons}, {Goldstein}, {Gommers},
  {Greco}, {Greenfield}, {Groener}, {Grollier}, {Hagen}, {Hirst}, {Homeier},
  {Horton}, {Hosseinzadeh}, {Hu}, {Hunkeler}, {Ivezi{\'c}}, {Jain}, {Jenness},
  {Kanarek}, {Kendrew}, {Kern}, {Kerzendorf}, {Khvalko}, {King}, {Kirkby},
  {Kulkarni}, {Kumar}, {Lee}, {Lenz}, {Littlefair}, {Ma}, {Macleod},
  {Mastropietro}, {McCully}, {Montagnac}, {Morris}, {Mueller}, {Mumford},
  {Muna}, {Murphy}, {Nelson}, {Nguyen}, {Ninan}, {N{\"o}the}, {Ogaz}, {Oh},
  {Parejko}, {Parley}, {Pascual}, {Patil}, {Patil}, {Plunkett}, {Prochaska},
  {Rastogi}, {Reddy Janga}, {Sabater}, {Sakurikar}, {Seifert}, {Sherbert},
  {Sherwood-Taylor}, {Shih}, {Sick}, {Silbiger}, {Singanamalla}, {Singer},
  {Sladen}, {Sooley}, {Sornarajah}, {Streicher}, {Teuben}, {Thomas},
  {Tremblay}, {Turner}, {Terr{\'o}n}, {van Kerkwijk}, {de la Vega}, {Watkins},
  {Weaver}, {Whitmore}, {Woillez}, {Zabalza}, \& {Contributors}}]{astropy:2018}
{Price-Whelan}, A.~M., {Sip{\H{o}}cz}, B.~M., {G{\"u}nther}, H.~M., {et~al.}
  2018, \aj, 156, 123

\bibitem[{{Quanz} {et~al.}(2015){Quanz}, {Amara}, {Meyer}, {Girard},
  {Kenworthy}, \& {Kasper}}]{Quanz2015}
{Quanz}, S.~P., {Amara}, A., {Meyer}, M.~R., {et~al.} 2015, \apj, 807, 64

\bibitem[{{Quanz} {et~al.}(2013){Quanz}, {Avenhaus}, {Buenzli}, {Garufi},
  {Schmid}, \& {Wolf}}]{Quanz2013}
{Quanz}, S.~P., {Avenhaus}, H., {Buenzli}, E., {et~al.} 2013, \apjl, 766, L2

\bibitem[{{Rameau} {et~al.}(2017){Rameau}, {Follette}, {Pueyo}, {Marois},
  {Macintosh}, {Millar-Blanchaer}, {Wang}, {Vega}, {Doyon}, {Lafreni{\`e}re},
  {Nielsen}, {Bailey}, {Chilcote}, {Close}, {Esposito}, {Males}, {Metchev},
  {Morzinski}, {Ruffio}, {Wolff}, {Ammons}, {Barman}, {Bulger}, {Cotten}, {De
  Rosa}, {Duchene}, {Fitzgerald}, {Goodsell}, {Graham}, {Greenbaum}, {Hibon},
  {Hung}, {Ingraham}, {Kalas}, {Konopacky}, {Larkin}, {Maire}, {Marchis},
  {Oppenheimer}, {Palmer}, {Patience}, {Perrin}, {Poyneer}, {Rajan},
  {Rantakyr{\"o}}, {Marley}, {Savransky}, {Schneider}, {Sivaramakrishnan},
  {Song}, {Soummer}, {Thomas}, {Wallace}, {Ward-Duong}, \&
  {Wiktorowicz}}]{Rameau2017}
{Rameau}, J., {Follette}, K.~B., {Pueyo}, L., {et~al.} 2017, \aj, 153, 244

\bibitem[{{Reggiani} {et~al.}(2018){Reggiani}, {Christiaens}, {Absil}, {Mawet},
  {Huby}, {Choquet}, {Gomez Gonzalez}, {Ruane}, {Femenia}, {Serabyn},
  {Matthews}, {Barraza}, {Carlomagno}, {Defr{\`e}re}, {Delacroix}, {Habraken},
  {Jolivet}, {Karlsson}, {Orban de Xivry}, {Piron}, {Surdej}, {Vargas Catalan},
  \& {Wertz}}]{Reggiani2018}
{Reggiani}, M., {Christiaens}, V., {Absil}, O., {et~al.} 2018, \aap, 611, A74

\bibitem[{{Reipurth} {et~al.}(1996){Reipurth}, {Pedrosa}, \&
  {Lago}}]{Reipurth1996}
{Reipurth}, B., {Pedrosa}, A., \& {Lago}, M.~T.~V.~T. 1996, \aaps, 120, 229

\bibitem[{{Riaud} \& {Schneider}(2007)}]{Riaud2007A&A...469..355R}
{Riaud}, P. \& {Schneider}, J. 2007, \aap, 469, 355

\bibitem[{{Rigliaco} {et~al.}(2012){Rigliaco}, {Natta}, {Testi}, {Randich},
  {Alcal{\`a}}, {Covino}, \& {Stelzer}}]{Rigliaco2012}
{Rigliaco}, E., {Natta}, A., {Testi}, L., {et~al.} 2012, \aap, 548, A56

\bibitem[{{Sallum} {et~al.}(2015){Sallum}, {Follette}, {Eisner}, {Close},
  {Hinz}, {Kratter}, {Males}, {Skemer}, {Macintosh}, {Tuthill}, {Bailey},
  {Defr{\`e}re}, {Morzinski}, {Rodigas}, {Spalding}, {Vaz}, \&
  {Weinberger}}]{Sallum2015}
{Sallum}, S., {Follette}, K.~B., {Eisner}, J.~A., {et~al.} 2015, \nat, 527, 342

\bibitem[{{Santamar{\'{\i}}a-Miranda}
  {et~al.}(2018){Santamar{\'{\i}}a-Miranda}, {C{\'a}ceres}, {Schreiber},
  {Hardy}, {Bayo}, {Parsons}, {Gromadzki}, \& {Aguayo
  Villegas}}]{Santamaria-Miranda2018}
{Santamar{\'{\i}}a-Miranda}, A., {C{\'a}ceres}, C., {Schreiber}, M.~R.,
  {et~al.} 2018, \mnras, 475, 2994

\bibitem[{{Schmid} {et~al.}(2018){Schmid}, {Bazzon}, {Roelfsema}, {Mouillet},
  {Milli}, {Menard}, {Gisler}, {Hunziker}, {Pragt}, {Dominik}, {Boccaletti},
  {Ginski}, {Abe}, {Antoniucci}, {Avenhaus}, {Baruffolo}, {Baudoz}, {Beuzit},
  {Carbillet}, {Chauvin}, {Claudi}, {Costille}, {Daban}, {de Haan}, {Desidera},
  {Dohlen}, {Downing}, {Elswijk}, {Engler}, {Feldt}, {Fusco}, {Girard},
  {Gratton}, {Hanenburg}, {Henning}, {Hubin}, {Joos}, {Kasper}, {Keller},
  {Langlois}, {Lagadec}, {Martinez}, {Mulder}, {Pavlov}, {Podio}, {Puget},
  {Quanz}, {Rigal}, {Salasnich}, {Sauvage}, {Schuil}, {Siebenmorgen}, {Sissa},
  {Snik}, {Suarez}, {Thalmann}, {Turatto}, {Udry}, {van Duin}, {van Holstein},
  {Vigan}, \& {Wildi}}]{Schmid2018}
{Schmid}, H.~M., {Bazzon}, A., {Roelfsema}, R., {et~al.} 2018, \aap, 619, A9

\bibitem[{{Sissa} {et~al.}(2018){Sissa}, {Gratton}, {Garufi}, {Rigliaco},
  {Zurlo}, {Mesa}, {Langlois}, {de Boer}, {Desidera}, {Ginski}, {Lagrange},
  {Maire}, {Vigan}, {Dima}, {Antichi}, {Baruffolo}, {Bazzon}, {Benisty},
  {Beuzit}, {Biller}, {Boccaletti}, {Bonavita}, {Bonnefoy}, {Brandner},
  {Bruno}, {Buenzli}, {Cascone}, {Chauvin}, {Cheetham}, {Claudi}, {Cudel}, {De
  Caprio}, {Dominik}, {Fantinel}, {Farisato}, {Feldt}, {Fontanive}, {Galicher},
  {Giro}, {Hagelberg}, {Incorvaia}, {Janson}, {Kasper}, {Keppler}, {Kopytova},
  {Lagadec}, {Lannier}, {Lazzoni}, {LeCoroller}, {Lessio}, {Ligi}, {Marzari},
  {Menard}, {Meyer}, {Mouillet}, {Peretti}, {Perrot}, {Potiron}, {Rouan},
  {Salasnich}, {Salter}, {Samland}, {Schmidt}, {Scuderi}, \&
  {Wildi}}]{Sissa2018}
{Sissa}, E., {Gratton}, R., {Garufi}, A., {et~al.} 2018, \aap, 619, A160

\bibitem[{{Snellen} {et~al.}(2015){Snellen}, {de Kok}, {Birkby}, {Brandl},
  {Brogi}, {Keller}, {Kenworthy}, {Schwarz}, \& {Stuik}}]{Snellen2015}
{Snellen}, I., {de Kok}, R., {Birkby}, J.~L., {et~al.} 2015, \aap, 576, A59

\bibitem[{{Soummer} {et~al.}(2012){Soummer}, {Pueyo}, \&
  {Larkin}}]{Soummer2012}
{Soummer}, R., {Pueyo}, L., \& {Larkin}, J. 2012, \apjl, 755, L28

\bibitem[{{Sparks} \& {Ford}(2002)}]{Sparks2002}
{Sparks}, W.~B. \& {Ford}, H.~C. 2002, \apj, 578, 543

\bibitem[{{Tannirkulam} {et~al.}(2008){Tannirkulam}, {Monnier}, {Harries},
  {Millan-Gabet}, {Zhu}, {Pedretti}, {Ireland}, {Tuthill}, {ten Brummelaar},
  {McAlister}, {Farrington}, {Goldfinger}, {Sturmann}, {Sturmann}, \&
  {Turner}}]{2008ApJ...689..513T}
{Tannirkulam}, A., {Monnier}, J.~D., {Harries}, T.~J., {et~al.} 2008, \apj,
  689, 513

\bibitem[{{Thalmann} {et~al.}(2016){Thalmann}, {Janson}, {Garufi},
  {Boccaletti}, {Quanz}, {Sissa}, {Gratton}, {Salter}, {Benisty}, {Bonnefoy},
  {Chauvin}, {Daemgen}, {Desidera}, {Dominik}, {Engler}, {Feldt}, {Henning},
  {Lagrange}, {Langlois}, {Lannier}, {Le Coroller}, {Ligi}, {M{\'e}nard},
  {Mesa}, {Meyer}, {Mulders}, {Olofsson}, {Pinte}, {Schmid}, {Vigan}, \&
  {Zurlo}}]{Thalmann2016}
{Thalmann}, C., {Janson}, M., {Garufi}, A., {et~al.} 2016, \apjl, 828, L17

\bibitem[{{Thalmann} {et~al.}(2015){Thalmann}, {Mulders}, {Janson}, {Olofsson},
  {Benisty}, {Avenhaus}, {Quanz}, {Schmid}, {Henning}, {Buenzli}, {M{\'e}nard},
  {Carson}, {Garufi}, {Messina}, {Dominik}, {Leisenring}, {Chauvin}, \&
  {Meyer}}]{Thalmann2015}
{Thalmann}, C., {Mulders}, G.~D., {Janson}, M., {et~al.} 2015, \apjl, 808, L41

\bibitem[{{Thanathibodee} {et~al.}(2019){Thanathibodee}, {Calvet}, {Bae},
  {Muzerolle}, \& {Hern{\'a}ndez}}]{Thanathibodee2019}
{Thanathibodee}, T., {Calvet}, N., {Bae}, J., {Muzerolle}, J., \&
  {Hern{\'a}ndez}, R.~F. 2019, \apj, 885, 94

\bibitem[{{van den Ancker} {et~al.}(1997){van den Ancker}, {The}, {Tjin A
  Djie}, {Catala}, {de Winter}, {Blondel}, \& {Waters}}]{vandenAncker1997}
{van den Ancker}, M.~E., {The}, P.~S., {Tjin A Djie}, H.~R.~E., {et~al.} 1997,
  \aap, 324, L33

\bibitem[{{van der Marel} {et~al.}(2019){van der Marel}, {Dong}, {di
  Francesco}, {Williams}, \& {Tobin}}]{vanderMarel}
{van der Marel}, N., {Dong}, R., {di Francesco}, J., {Williams}, J.~P., \&
  {Tobin}, J. 2019, \apj, 872, 112

\bibitem[{{van Terwisga} {et~al.}(2018){van Terwisga}, {van Dishoeck},
  {Ansdell}, {van der Marel}, {Testi}, {Williams}, {Facchini}, {Tazzari},
  {Hogerheijde}, {Trapman}, {Manara}, {Miotello}, {Maud}, \&
  {Harsono}}]{vanTerwisga2018}
{van Terwisga}, S.~E., {van Dishoeck}, E.~F., {Ansdell}, M., {et~al.} 2018,
  \aap, 616, A88

\bibitem[{{Vioque} {et~al.}(2018){Vioque}, {Oudmaijer}, {Baines},
  {Mendigut{\'\i}a}, \& {P{\'e}rez-Mart{\'\i}nez}}]{Vioque2018}
{Vioque}, M., {Oudmaijer}, R.~D., {Baines}, D., {Mendigut{\'\i}a}, I., \&
  {P{\'e}rez-Mart{\'\i}nez}, R. 2018, \aap, 620, A128

\bibitem[{{Wagner} {et~al.}(2018){Wagner}, {Follete}, {Close}, {Apai}, {Gibbs},
  {Keppler}, {M{\"u}ller}, {Henning}, {Kasper}, {Wu}, {Long}, {Males},
  {Morzinski}, \& {McClure}}]{Wagner2018}
{Wagner}, K., {Follete}, K.~B., {Close}, L.~M., {et~al.} 2018, \apjl, 863, L8

\bibitem[{{Wagner} {et~al.}(2019){Wagner}, {Stone}, {Spalding}, {Apai}, {Dong},
  {Ertel}, {Leisenring}, \& {Webster}}]{Wagner2019}
{Wagner}, K., {Stone}, J.~M., {Spalding}, E., {et~al.} 2019, \apj, 882, 20

\bibitem[{{Wang} {et~al.}(2017){Wang}, {Mawet}, {Ruane}, {Hu}, \&
  {Benneke}}]{Wang2017}
{Wang}, J., {Mawet}, D., {Ruane}, G., {Hu}, R., \& {Benneke}, B. 2017, \aj,
  153, 183

\bibitem[{{Weilbacher} {et~al.}(2015){Weilbacher}, {Monreal-Ibero},
  {Kollatschny}, {Ginsburg}, {McLeod}, {Kamann}, {Sandin}, {Palsa}, {Wisotzki},
  {Bacon}, {Selman}, {Brinchmann}, {Caruana}, {Kelz}, {Martinsson},
  {P{\'e}contal-Rousset}, {Richard}, \& {Wendt}}]{Weilbacher2015}
{Weilbacher}, P.~M., {Monreal-Ibero}, A., {Kollatschny}, W., {et~al.} 2015,
  \aap, 582, A114

\bibitem[{{Weilbacher} {et~al.}(2020){Weilbacher}, {Palsa}, {Streicher},
  {Bacon}, {Urrutia}, {Wisotzki}, {Conseil}, {Husemann}, {Jarno}, {Kelz},
  {P{\'e}contal-Rousset}, {Richard}, {Roth}, {Selman}, \&
  {Vernet}}]{Weilbacher2020}
{Weilbacher}, P.~M., {Palsa}, R., {Streicher}, O., {et~al.} 2020, \aap, 641,
  A28

\bibitem[{{Wisotzki} {et~al.}(2016){Wisotzki}, {Bacon}, {Blaizot},
  {Brinchmann}, {Herenz}, {Schaye}, {Bouch{\'e}}, {Cantalupo}, {Contini},
  {Carollo}, {Caruana}, {Courbot}, {Emsellem}, {Kamann}, {Kerutt}, {Leclercq},
  {Lilly}, {Patr{\'\i}cio}, {Sandin}, {Steinmetz}, {Straka}, {Urrutia},
  {Verhamme}, {Weilbacher}, \& {Wendt}}]{2016A&A...587A..98W}
{Wisotzki}, L., {Bacon}, R., {Blaizot}, J., {et~al.} 2016, \aap, 587, A98

\bibitem[{{Wisotzki} {et~al.}(2018){Wisotzki}, {Bacon}, {Brinchmann},
  {Cantalupo}, {Richter}, {Schaye}, {Schmidt}, {Urrutia}, {Weilbacher},
  {Akhlaghi}, {Bouch{\'e}}, {Contini}, {Guiderdoni}, {Herenz}, {Inami},
  {Kerutt}, {Leclercq}, {Marino}, {Maseda}, {Monreal-Ibero}, {Nanayakkara},
  {Richard}, {Saust}, {Steinmetz}, \& {Wendt}}]{2018Natur.562..229W}
{Wisotzki}, L., {Bacon}, R., {Brinchmann}, J., {et~al.} 2018, \nat, 562, 229

\bibitem[{{Zacharias} {et~al.}(2012){Zacharias}, {Finch}, {Girard}, {Henden},
  {Bartlett}, {Monet}, \& {Zacharias}}]{Zacharias+2012}
{Zacharias}, N., {Finch}, C.~T., {Girard}, T.~M., {et~al.} 2012, VizieR Online
  Data Catalog, I/322A

\bibitem[{{Zhang} {et~al.}(2015){Zhang}, {Blake}, \& {Bergin}}]{Zhang2015}
{Zhang}, K., {Blake}, G.~A., \& {Bergin}, E.~A. 2015, \apjl, 806, L7

\bibitem[{{Zhu}(2015)}]{Zhu2015}
{Zhu}, Z. 2015, \apj, 799, 16

\bibitem[{{Zhu} {et~al.}(2014){Zhu}, {Stone}, {Rafikov}, \& {Bai}}]{Zhu2014}
{Zhu}, Z., {Stone}, J.~M., {Rafikov}, R.~R., \& {Bai}, X.-n. 2014, \apj, 785,
  122

\bibitem[{{Zurlo} {et~al.}(2020){Zurlo}, {Cugno}, {Montesinos}, {Perez},
  {Canovas}, {Casassus}, {Christiaens}, {Cieza}, \& {Huelamo}}]{Zurlo2020}
{Zurlo}, A., {Cugno}, G., {Montesinos}, M., {et~al.} 2020, \aap, 633, A119

\end{thebibliography}

\begin{appendix}
%\cite{Weilbacher2015}

\section{Blob ghost}
\label{sec:blob_ghost}

\begin{figure}
   \centering
    \includegraphics[width=0.48\textwidth]{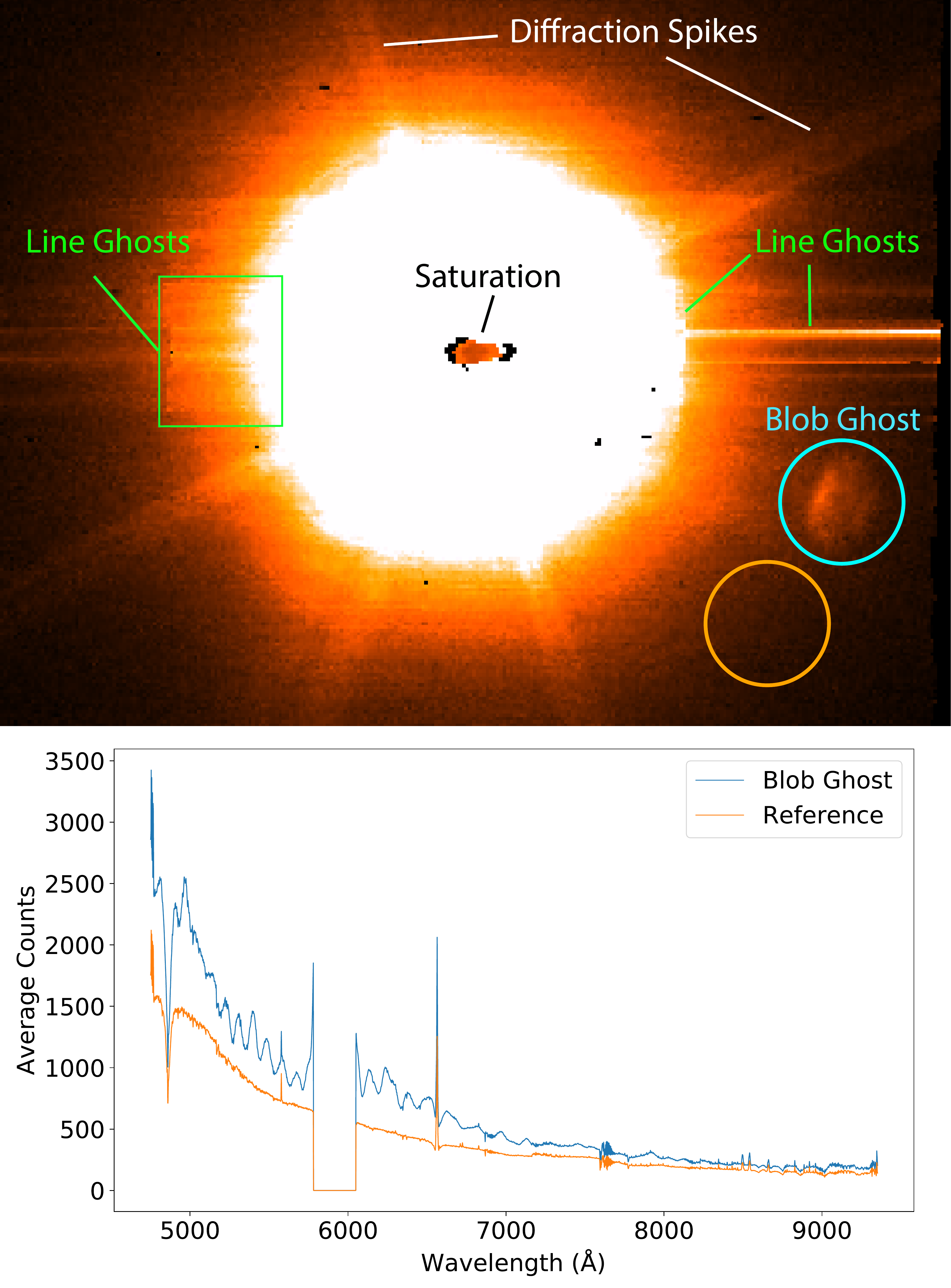}
      \caption{Top panel:  Image of HD 163296 at 6560 \AA, showing multiple instrumental artifacts. Line ghosts are clearly visible as overbright strips in the entire horizontal IFU and a few slices. The blob ghost is shown as  the cyan circle. The spectra of the blob ghost and nearby reference (orange circle) are shown in the bottom panel. 
%      Diffraction spikes and saturation are shown marked. 
      }
    \label{fig:blob_ghost}
\end{figure}

\cite{Weilbacher2015} identified four artifacts visible in the MUSE wide field mode (WFM) observation as saturation, ghosts, diffraction spikes, and second spectral order. In this work we find a new kind of ghost that has a blob morphology with spectral fringing patterns that  strongly contaminate the blue part of the spectrum. Figure~\ref{fig:blob_ghost}   identifies the  different kinds of known instrumental artifacts in MUSE and makes  a spectral comparison between the nearby sky reference and the blob ghost. Diffraction spikes can be removed in HRSDI and saturation can be avoided by optimizing the exposure time. Since NFM observations are done in the normal mode with a blocking filter, second-order contamination is blocked.

%In Figure~\ref{jets_cnt}, we show the jets in HD163296. 
The arc-like structure near the SW jet (see Fig.~\ref{jets_cnt}) is at the same location as the blob ghost in the raw image of HD 163296 before the removal of stellar emission (see Fig.~\ref{fig:blob_ghost}). %The dataset we show has an exposure setting of 250s. 
In the dataset with the shorter integration time, we do not find any blob ghosts at the same or other locations, suggesting their non-astronomical origin. The instrumental origin of the blob ghost is not fully understood, but it is likely from the laser-guided AO system and the internal reflection, which gives a wavelength-dependent fringing pattern. %affecting mostly on the blue part of spectra. 
We note that line ghosts do not have periodic spectral fringing, suggesting a different instrumental origin.

Blob ghosts can also be removed if necessary. The signal frequency of the blob ghost is different from stellar line emission (i.e., the \Ha~line in the bottom panel in Fig.~\ref{fig:blob_ghost}). Therefore, applying a band-stop filter before PCA can remove the spectral fringing caused by blob ghosts, while the emission lines from planets and jets are preserved. All differential broadband features will be aligned and removed during the low-order correction in HRSDI.

\end{appendix}

\end{document}